\documentclass[aip,jcp,reprint,superscriptaddress,noshowkeys]{revtex4-2}
\usepackage{graphicx,dcolumn,bm,microtype,multirow,amscd,amsmath,amssymb,amsfonts,physics,wrapfig,siunitx}

\usepackage[version=4]{mhchem}
\usepackage[dvipsnames]{xcolor}
\usepackage[normalem]{ulem}
\usepackage[utf8]{inputenc}

\usepackage{hyperref}
\hypersetup{
    colorlinks,
    linkcolor={red!50!black},
    citecolor={red!70!black},
    urlcolor={red!80!black}
}

\usepackage[normalem]{ulem}

\newcommand{\titou}[1]{\textcolor{black}{#1}}

\newcommand{\LCPQ}{Laboratoire de Chimie et Physique Quantiques (UMR 5626), Universit\'e de Toulouse, CNRS, Toulouse, France}
\newcommand{\LPT}{Laboratoire de Physique Th\'eorique, Universit\'e de Toulouse, CNRS, and European Theoretical Spectroscopy Facility (ETSF), Toulouse, France}
\newcommand{\NEEL}{Universit\'e Grenoble Alpes, CNRS, Institut N\'EEL, F-38042 Grenoble, France}

\begin{document}	

\title{Anomalous propagators and the particle-particle channel: Bethe-Salpeter equation}

\author{Antoine \surname{Marie}}
	\email{amarie@irsamc.ups-tlse.fr}
	\affiliation{\LCPQ}
\author{Pina \surname{Romaniello}}
	\affiliation{\LPT}
\author{Xavier \surname{Blase}}
        \affiliation{\NEEL}
\author{Pierre-Fran\c{c}ois \surname{Loos}}
	\email{loos@irsamc.ups-tlse.fr}
	\affiliation{\LCPQ}

\begin{abstract}
  The Bethe-Salpeter equation has been extensively employed to compute the two-body electron-hole propagator and its poles, which correspond to the neutral excitation energies of the system.
  Through a different time-ordering, the two-body Green's function can also describe the propagation of two electrons or two holes.
  The corresponding poles are the double ionization potentials and double electron affinities of the system.
  In this work, a Bethe-Salpeter equation for the two-body particle-particle propagator is derived within the linear-response formalism using a pairing field and anomalous propagators.
  This framework allows us to compute kernels corresponding to different self-energy approximations ($GW$, $T$-matrix, and second-Born) as in the usual electron-hole case.
  The performance of these various kernels is gauged for singlet and triplet valence double ionization potentials using a set of 23 small molecules.
  The description of double core hole states is also analyzed.
\end{abstract}

\maketitle

\section{Introduction}

Despite its conceptual simplicity, the electron-hole (eh) random-phase approximation (RPA),  introduced by Bohm and Pines in the context of the uniform electron gas, \cite{Bohm_1951,Pines_1952,Bohm_1953,Nozieres_1958,Gell-Mann_1957} has proven effective in describing a variety of physical phenomena.\footnote{The aforementioned RPA formalism is usually referred to as particle-hole (ph) RPA. However, for notational consistency, we prefer the terminology eh-RPA in the present context.}
These include collective excitations (plasmons) in metals and semiconductors, \cite{Martin_2016} non-covalent interaction energies, \cite{Lu_2009,Zhu_2010,Nguyen_2020} and screening of the Coulomb interaction due to density fluctuations. \cite{Martin_2016,Reining_2018,Golze_2019,Marie_2024a}
However, eh-RPA falls short in accurately describing molecular excitations and excitons in insulators, \cite{Benedict_1998,Benedict_1998a,Rohlfing_1998,Rohlfing_2000,Martin_2016} highlighting the need for methods that go beyond eh-RPA.

Since its inception, the eh-RPA equations have been derived through various formalisms, such as Rowe's equation of motion (EOM), \cite{Rowe_1968,Rowe_1968a,Mertins_1996,Simons_2005} time-dependent density-functional theory (TDDFT), \cite{Runge_1984,UlrichBook} and the Bethe-Salpeter equation (BSE). \cite{Salpeter_1951,Strinati_1988,Blase_2018,Blase_2020}
Each of these frameworks offers a pathway to devising approximations that extend beyond RPA.
Amongst these, TDDFT has been by far the most popular, though it suffers from well-documented drawbacks. \cite{Casida_1998,Dreuw_2003,Peach_2011,LeGuennic_2015}
In particular, TDDFT lacks systematic improvement, as progressing up Jacob's ladder of exchange-correlation functionals \cite{Perdew_2001} does not guarantee higher accuracy (see Ref.~\onlinecite{Teale_2022} for a recent review).
In contrast, EOM can be systematically improved by increasing the rank of the excitation operator or moving beyond a mean-field reference. \cite{Mertins_1996,Simons_2005}
For example, EOM coupled-cluster (EOM-CC) is the method of choice for highly accurate excitation energies, albeit at a significant computational cost. \cite{Hirata_2000,Kallay_2004b,Matthews_2015,Loos_2020d,Loos_2021a}
 
Within the BSE formalism, eh-RPA corresponds to computing the two-body eh propagator $L$, known as the polarization propagator, using the simplest approximation of the kernel, namely, the Coulomb kernel.
To go beyond eh-RPA, one must improve the kernel using, for example, perturbation theory or Hedin's equations. \cite{Martin_2016}
For example, the kernel derived from the popular $GW$ self-energy approximation \cite{Martin_1959,Reining_2018,Golze_2019,Marie_2024a} has been highly successful in computing low-lying excited states of various natures (charge transfer, Rydberg, valence, etc) in molecular systems with a very attractive accuracy/cost ratio. \cite{vanderHorst_1999,Rohlfing_1999,Rohlfing_2000,Boulanger_2014,Bruneval_2015,Jacquemin_2015a,Hirose_2015,Jacquemin_2017a,Jacquemin_2017b,Gui_2018,Blase_2018,Blase_2020,Liu_2020,Loos_2021b,Knysh_2024}
Perturbative kernels based on expansions in the Coulomb interaction, or in alternative effective interactions like the $T$-matrix, have also been explored. \cite{Zhang_2013,Rebolini_2016,Loos_2022a,Orlando_2023a,Monino_2023}

\begin{figure}
  \centering
  \includegraphics[width=0.98\linewidth]{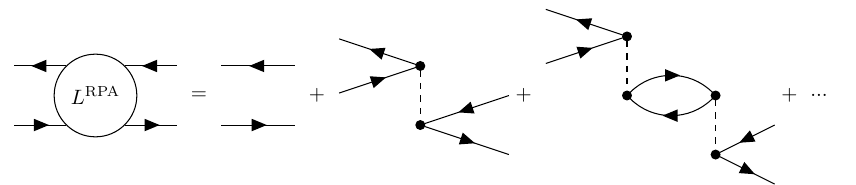}
  \includegraphics[width=0.98\linewidth]{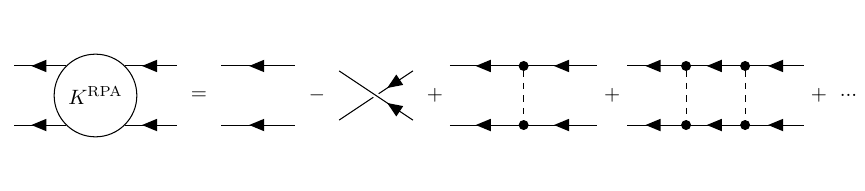}
  \caption{Diagrammatic representation of the eh propagator $L$ (top) and pp propagator $K$ (bottom) at the RPA level.
  The dashed lines represent the Coulomb interaction and the solid lines with arrows denote the one-body propagator.
  The first and second-order exchange terms have not been represented but should be included in $K^\RPA$.}
   \label{fig:rpa}
\end{figure}

The eh-RPA possesses an analog known as particle-particle (pp) RPA or pairing vibration approximation. \cite{Schuck_Book}
This method provides the simplest beyond-independent-particle approximation to compute the two-body pp propagator $K$. \cite{Fukuda_1964,Martin_2016}
Their close relationship is elegantly illustrated in the diagrammatic language of many-body perturbation theory, as depicted in Fig.~\Ref{fig:rpa}. \cite{MattuckBook}
Both approximations involve the resummation of specific classes of diagrams: bubbles for eh-RPA and pp ladders for pp-RPA.
The pp propagator captures complementary information to the polarization propagator as its poles correspond to the double ionization potentials (DIPs) and double electron affinities (DEAs) of the $N$-electron system.

In a recent series of papers, Weitao Yang's group has shown that pp-RPA offers an alternative approach to accessing excited states in $N$-electron systems. \cite{Yang_2013b,Yang_2014a,Yang_2015,Yang_2016,Yang_2017a,Li_2024a,Li_2024b}
Specifically, excitation energies of the $N$-electron system can be computed as the difference between DEAs from the corresponding ($N-2$)-electron reference state.
(For a complementary starting point using the ($N+2$)-electron system, see Refs.~\onlinecite{Bannwarth_2020,Li_2024a}.) 
Strategies along these lines have already been applied within coupled-cluster (CC) theory \cite{Sattelmeyer_2003,Musial_2011,Musial_2012,Shen_2013,Musial_2014,Perera_2016,Gulania_2021} and is also related to the spin-flip method for excited states. \cite{Krylov_2001a,Krylov_2001b,Shao_2003,Krylov_2006,Bernard_2012,Lefrancois_2015,Monino_2021}
Unlike linear-response-based methods, this approach avoids bias toward the ground state because the ground and excited states are obtained within the same calculation.
Yang and co-workers have shown that pp-RPA has some advantages over eh-RPA and TDDFT, such as the ability to describe charge transfer, \cite{Yang_2013b,Yang_2017a} doubly excited states, \cite{Yang_2013b} or conical intersections. \cite{Yang_2016}
However, the accuracy of pp-RPA is not always satisfactory, motivating the development of methods that go beyond pp-RPA.

As with eh-RPA, pp-RPA can be derived and extended through various theoretical frameworks. 
The EOM in the $(N+2)$- and ($N-2)$-electron sectors of Fock space leads to pp-RPA and, as for the $N$-electron case, the systematically-improvable DIP-EOM-CC and DEA-EOM-CC formalisms can provide reliable reference energies. \cite{Musial_2011,Musial_2012,Shen_2013,Gururangan_2025}
On the other hand, improving pp-RPA within the pairing-field TDDFT framework is even more challenging than in standard TDDFT. \cite{vanAggelen_2013,vanAggelen_2014}
This difficulty arises because the pairing-field TDDFT kernel is obtained by differentiating the exchange-correlation functional with respect to the anomalous density. However, the number of functionals for the anomalous density remains quite limited \cite{Oliveira_1988b,Luders_2005,Marques_2005} and, to the best of our knowledge, these functionals have not been employed to derive kernels within pairing-field TDDFT.
Finally, the pp channel of the BSE has been much less explored and developed than its eh counterpart.
Although the BSE for the pp propagator has been reported in the literature, its kernel lacks a simple form. 
In fact, it is expressed as a Dyson equation of the eh-BSE kernel, making it difficult to apply in practice. \cite{CsanakBook}

The main result of this paper is to present an alternative expression of the pp-BSE kernel by considering pairing fields and anomalous Green's functions.
This new expression is fully analog to the eh-BSE kernel as both are expressed as the derivative of a self-energy with respect to a propagator.
The derivation of this expression is detailed in Sec.~\ref{sec:bse} after discussing correlation functions and their link to response functions in Sec.~\ref{sec:corr_funct} and \ref{sec:resp_funct}, respectively.
The last subsection of Sec.~\ref{sec:theoretical} addresses the finite orbital basis formulation of the pp-BSE.
Then, the pp-BSE analogs of the usual eh-BSE kernels [GF(2), \cite{Suhai_1983,Casida_1989,Casida_1991,Hirata_2015} $GW$, \cite{Martin_2016,Reining_2018,Golze_2019,Marie_2024a} and pp $T$-matrix \cite{Bethe_1957,Baym_1961,Danielewicz_1984a,Zhang_2017,Orlando_2023b}] are derived in Sec.~\ref{sec:kernels} and their specificities are discussed.
Section \ref{sec:comp_det} reports the computational details of our implementation, while our numerical results are presented in Sec.~\ref{sec:results}.
Some concluding remarks are provided in Sec.~\ref{sec:conclusion}.

\section{Theoretical framework}
\label{sec:theoretical}

\begin{figure*}
  \centering
  \includegraphics[width=0.8\linewidth]{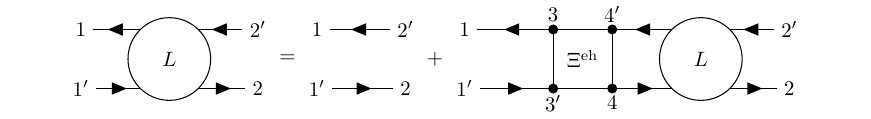}
  \caption{Diagrammatic representation of the eh-BSE, as defined in Eq.~\eqref{eq:ehBSE}.}
  \label{fig:ehBSE}
\end{figure*}

\subsection{Correlation functions}
\label{sec:corr_funct}

The equilibrium time-ordered two-body Green's function (at zero temperature) is defined as
\begin{equation}
  \label{eq:G2}
  G_2(12;1'2') = (-\ii)^2\mel{\Psi_0^N}{\hT[\hpsi(1)\hpsi(2)\hpsid(2')\hpsid(1')]}{\Psi_0^N},
\end{equation}
where $1$ is a space-spin-time composite variable $(1)=(\vb{x}_1,t_1)=(\vb{r}_1,\sigma_1,t_1)$, $\Hat{T}$ is the time-ordering operator and $\ket*{\Psi_0^N}$ is the exact $N$-electron ground state.
The annihilation and creation field operators are in the Heisenberg representation, that is,
$\hpsi(1) = e^{i\hH t_1}\Hat{\psi}(\vb{x}_1)e^{-i\hH t_1}$,
where $\hH$ is the non-relativistic electronic Hamiltonian
\begin{multline}
  \label{eq:hamiltonian}
  \hH = \int \dd{(\bx_1\bx_{1'})} \hpsid(\bx_1) h(\bx_1\bx_{1'}) \hpsi(\bx_{1'}) 
  \\
  + \frac{1}{2} \iint \dd{(\bx_1\bx_2\bx_{1'}\bx_{2'})} 
  \\
  \times \hpsid(\bx_1) \hpsid(\bx_2) v(\bx_1\bx_2;\bx_{1'}\bx_{2'}) \hpsi(\bx_{2'}) \hpsi(\bx_{1'}).
\end{multline}
The 4-point Coulomb interaction is defined as
\begin{equation}
  \label{eq:4point_coulomb}
  v(\bx_1\bx_2;\bx_{1'}\bx_{2'}) = \frac{\delta(\bx_1\bx_{1'})\delta(\bx_2\bx_{2'})}{\abs{\br_1-\br_2}}
\end{equation}
and $h(\bx_1\bx_{1'})$ is the one-body Hamiltonian.

The two-body Green's function describes the propagation of two particles.
These particles can be either electrons or holes depending on the ordering of the time variables.
For example, if $t_2,t_{2'}>t_1,t_{1'}$, then $G_2$ reduces to
\begin{multline}
  \label{eq:ehG2}
  G_2(12;1'2') 
  \\ 
  = (-\ii)^2 \mel{\Psi_0^N}{\hT\mqty[\hpsi(2)\hpsid(2')]\hT\mqty[\hpsi(1)\hpsid(1')]}{\Psi_0^N}
\end{multline}
and describes the propagation of an electron-hole pair.
The eh correlation function $L$ associated with $G_2$ is obtained by removing the uncorrelated part of Eq.~\eqref{eq:ehG2}, that is,
\begin{equation}
  \label{eq:ehL}
  L(12;1'2') = -G_2(12;1'2') + G(11')G(22'),
\end{equation}
where we have introduced the one-body Green's function
\begin{equation}
  \label{eq:G1}
  G(11') = (-\mathrm{i})\mel{\Psi_0^N}{\hT\mqty[\hpsi(1)\hpsid(1')]}{\Psi_0^N},
\end{equation}
which describes the propagation of either a hole or an electron.

One can further assume that the electron and hole forming the pair are created and annihilated simultaneously by setting $t_2 = t_{2'}^+$ and $t_1 = t_{1'}^+$ (where $t^+ = t + \eta$ with $\eta$ an infinitesimally small positive shift) such that $L(12;1'2')$ depends only on the time difference $t_2 - t_1$.
The frequency-space representation of this correlation function is
  \begin{equation}
    \begin{split}
      L(\bx_1\bx_2;\bx_{1'}\bx_{2'};\omega) 
      & = \sum_{n > 0} \frac{L_{n}^N(\bx_2\bx_{2'})R_{n}^N(\bx_1\bx_{1'})}{\omega - (E_n^N - E_0^N - \ii\eta)}
      \\
      & - \sum_{n > 0} \frac{L_{n}^N(\bx_2\bx_{2'})R_{n}^N(\bx_1\bx_{1'})}{\omega - (E_0^N - E_n^N + \ii\eta)},
    \end{split}
  \end{equation}
with amplitudes
\begin{subequations}
\begin{align}
  L_{n}^N(\bx_1\bx_{1'}) & = \mel{\Psi_0^N}{\hpsid(\bx_1)\hpsi(\bx_{1'})}{\Psi_n^N},
  \\
  R_{n}^N(\bx_1\bx_{1'}) & = \mel{\Psi_n^N}{\hpsid(\bx_1)\hpsi(\bx_{1'})}{\Psi_0^N},
\end{align}
\end{subequations}
where $E_n^N$ and $\ket*{\Psi_n^N}$ are the energy and wave function of the $n$th excited state ($n = 0$ being the ground state) of the $N$-electron system.
This representation evidences that $L$ is directly linked to the excitation energies of the $N$-electron system.

If instead we impose the following time ordering, $t_1,t_2>t_{1'},t_{2'}$, then $G_2$ becomes
\begin{multline}
  \label{eq:hhG2}
  G_2(12;1'2') 
  \\
  = (-\ii)^2\mel{\Psi_0^N}{\hT\mqty[\hpsi(1)\hpsi(2)]\hT\mqty[\hpsid(2')\hpsid(1')]}{\Psi_0^N},
\end{multline}
and describes the propagation of two electrons.
Similarly, the time ordering $t_{1'},t_{2'}>t_1,t_2$ would describe the propagation of two holes.
In this case, one defines \footnote{The factor of 2 in Eq.\eqref{eq:hhL} is introduced to ensure that the definition of $K$ as a correlation function aligns with its formulation as a linear-response quantity in Eq.\eqref{eq:pp_schwinger}.} the associated correlation function as \cite{Martin_2016,Marie_2024b} 
\begin{equation}
  \label{eq:hhL}
  -2 K(12;1'2') = -G_2(12;1'2') + G^\hh(12)G^\ee(2'1'),
\end{equation}
where the uncorrelated part involves the hole-hole (hh) and electron-electron (ee) anomalous propagators
\begin{subequations}
  \begin{align}
    \label{eq:G1_hh}
    G^\hh(12)   &= (-\mathrm{i})\mel{\Psi_0^N}{\Hat{T}\mqty[ \hpsi(1)   \hpsi(2) ]}{\Psi_0^N}, \\
    \label{eq:G1_ee}
    G^\ee(2'1') &= (-\mathrm{i})\mel{\Psi_0^N}{\Hat{T}\mqty[\hpsid(2') \hpsid(1')]}{\Psi_0^N}.
  \end{align}
\end{subequations}
Note that, because the wave function $\ket*{\Psi_0^N}$ has a fixed number of particles, the uncorrelated part vanishes in this case.

Here again, the pairs are assumed to be created and annihilated instantaneously, \ie, {$t_1 = t_{2}^+$ and $t_{1'} = t_{2'}^+$}.
The Fourier transform of the pp propagator with respect to $t_1 - t_{1'}$ yields
\begin{equation}
  \label{eq:lehman_K}
  \begin{split}
    K(\bx_1\bx_2;\bx_{1'}\bx_{2'};\omega) 
    & = \frac{1}{2} \sum_{n} \frac{L_n^{N+2}(\bx_1\bx_2)R_n^{N+2}(\bx_{2'}\bx_{1'})}{\omega - (E_n^{N+2} - E_0^N - \ii\eta)}
    \\
    & - \frac{1}{2} \sum_{n} \frac{L_n^{N-2}(\bx_{2'}\bx_{1'})R_n^{N-2}(\bx_1\bx_2)}{\omega - (E_0^N - E_n^{N-2} + \ii\eta)},
  \end{split}
\end{equation}
with amplitudes
\begin{subequations}
  \begin{align}
    L_n^{N+2}(\bx_1\bx_2) & = \mel{\Psi_0^N}{\hpsi(\bx_1)\hpsi(\bx_2)}{\Psi_n^{N+2}}, 
    \\
    R_n^{N+2}(\bx_1\bx_2) & = \mel{\Psi_n^{N+2}}{\hpsid(\bx_{1})\hpsid(\bx_{2})}{\Psi_0^N},
    \\
    L_n^{N-2}(\bx_1\bx_2) & = \mel{\Psi_0^N}{\hpsid(\bx_{1})\hpsid(\bx_{2})}{\Psi_n^{N-2}},
    \\
    R_n^{N-2}(\bx_1\bx_2) & = \mel{\Psi_n^{N-2}}{\hpsi(\bx_1)\hpsi(\bx_2)}{\Psi_0^N}, 
  \end{align}
\end{subequations}
In the previous expressions, $E_n^{N\pm2}$ and $\ket*{\Psi_n^{N\pm2}}$ are the energy and wave function of the $n$th excited state of the $(N\pm2)$-electron system.
This highlights the direct link between the pp correlation function and the DIPs and DEAs of the $N$-electron system.

\begin{figure*}
  \centering
  \includegraphics[width=0.8\linewidth]{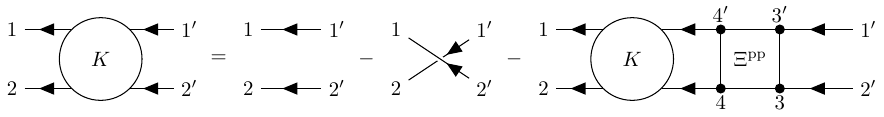}
  \caption{Diagrammatic representation of the pp-BSE [see Eq.~\eqref{eq:ppBSE}]. 
  The rightmost $K_0$ has been replaced by its first term [see Eq.~\eqref{eq:K0}] using the antisymmetry of the kernel.}
  \label{fig:ppBSE}
\end{figure*}

\subsection{Response functions}
\label{sec:resp_funct}

In the previous section, $L$ has been introduced as a correlation function but it can alternatively be regarded as a generalized response function. \cite{Martin_2016}
This is evidenced by the Schwinger relation \cite{Martin_1959,Strinati_1988}
\begin{equation}
  \label{eq:eh_schwinger}
  L(12;1'2') = \eval{\fdv{G(11';[U])}{U^\eh(2'2)}}_{U=0},
\end{equation}
where the time-dependent external potential
\begin{equation}
  \label{eq:eh_potential}
  \hat{\mathcal{U}}^\eh(t) = \int \dd(\bx_1\bx_{1'}) \hpsid(\bx_1) U^\eh(\bx_1\bx_{1'};t) \hpsi(\bx_{1'})
\end{equation}
has been added to the Hamiltonian \eqref{eq:hamiltonian} and $U^\eh(11') = U^\eh(\bx_{1}\bx_{1'};t_1)\delta(t_{1} - t_{1'})$.
The notation $G(11';[U])$ means that the propagator is computed in the presence of the external potential.
However, this dependence is not explicitly written in the following for the sake of conciseness.
The Schwinger relation shows that knowing the eh correlation function is equivalent to knowing the response of the one-body Green's function with respect to an external potential.

The pp correlation function can also be expressed as a functional derivative within the linear-response formalism.
This is done by considering the response of an anomalous Green's function to a time-dependent pairing potential
\begin{equation}
  \label{eq:pp_potential}
\begin{split}
  \hat{\mathcal{U}}^\pp(t) 
  = \frac{1}{2} \Bigg[ & \int \dd{(\bx_1\bx_{1'})} \hpsi(\bx_1) U^\hh(\bx_1\bx_{1'};t) \hpsi(\bx_{1'}) 
  \\
   + & \int \dd{(\bx_1\bx_{1'})} \hpsid(\bx_{1'}) U^\ee(\bx_1\bx_{1'};t) \hpsid(\bx_{1'}) \Bigg].
\end{split}
\end{equation}
Note that, in the presence of this perturbation, the Hamiltonian does not commute with the particle number operator.
Therefore, in the presence of the pairing potential, the wave function breaks the particle-number symmetry, and the anomalous Green's functions do not vanish anymore.
Following a derivation similar to the one of Eq.~\eqref{eq:eh_schwinger} (see \SupInf of Ref.~\onlinecite{Marie_2024b}), one can obtain \cite{Bickers_2004,PhDZhang}
\begin{equation}
  \label{eq:pp_schwinger}
  K(12;1'2') = \eval{\fdv{G^\ee(2'1')}{U^\hh(12)}}_{U=0},
\end{equation}
with $U^\hh(12) = U^\hh(\bx_{1}\bx_{2};t_1)\delta(t_{1} - t_{2})$.
This expression shows that the response of an anomalous propagator to a pairing field (evaluated at $U=0$) is non-zero even if the equilibrium Hamiltonian preserves the number of particles.
Therefore, through their derivatives, anomalous quantities can be useful even in a number-conserving framework.
This was already highlighted in pairing-field TDDFT, \cite{vanAggelen_2013,vanAggelen_2014} and will be further evidenced by the pp-BSE.

\subsection{Bethe-Salpeter equations}
\label{sec:bse}

The usual eh-BSE will be briefly reviewed before discussing in more depth its pp counterpart.
The eh-BSE is derived starting from the response-function form of $L$ [see Eq.~\eqref{eq:eh_schwinger}] and the Dyson equation for the one-body Green's function
\begin{equation}
  \label{eq:inverse_dyson}
  G^{-1}(12) = G_0^{-1}(12) - U^\eh(12) - \Sigma(12),
\end{equation}
where $G_0$ is the non-interacting one-body Green's function and $\Sigma$ is the exact self-energy containing Hartree, exchange, and correlation effects. \cite{Martin_2016}
This leads to a Dyson equation for $L$, which reads
\begin{multline}
  \label{eq:ehBSE}
  L(12;1'2') = L_0(12;1'2') \\
  + \int \dd(343'4') L_0(13';1'3) \Xi^\eh(34;3'4') L(4'2;42'), \\
\end{multline}
where $L_0(12;1'2') = G(12')G(21')$ is the non-interacting eh propagator and 
\begin{equation}
  \label{eq:eh_kernel}
  \Xi^\eh(34;3'4') = \fdv{\Sigma(33')}{G(4'4)}
\end{equation}
is the eh effective interaction kernel.
This equation is represented diagrammatically in Fig.~\ref{fig:ehBSE}.
Equation \eqref{eq:ehBSE} shows that the two-body propagator depends on $L_0$ and $\Xi^\eh$.
Hence, in practice, two approximations have to be made, namely the choice of the one-body Green's function that enters in $L_0$ and the self-energy considered to compute $\Xi^\eh$.

Now that the pp propagator has also been expressed in a response-function form [see Eq.~\eqref{eq:pp_schwinger}], a similar derivation can be performed for the pp channel.
Below, only the main steps of the derivation of the pp-BSE are discussed but a fully detailed derivation is provided in the \SupInf.
As a preliminary step, one introduces the Gorkov propagator \cite{Gorkov_1958}
\begin{equation}
  \label{eq:gorkov_propag}
  \bG(11') = \mqty(G^\he(11') & G^\hh(11') \\ G^\ee(11') & G^\eh(11')),
\end{equation}
which gathers the normal one-body propagator $G^\he \equiv G$ [see Eq.~\eqref{eq:G1}], the anomalous propagators [see Eqs.~\eqref{eq:G1_hh} and \eqref{eq:G1_ee}], as well as the eh one-body propagator related to the normal propagator as $G^\eh(11') = -G^\he(1'1)$.
This matrix formalism is known as Nambu's formalism. \cite{Nambu_1960}
The derivation of the pp-BSE then starts from Eq.~\eqref{eq:pp_schwinger} and relies on the following relation \cite{Marie_2024b}
\begin{equation}
  \label{eq:inverse_derivative_anom}
  \eval{\fdv{G^\ee(2'1')}{U^\hh(12)}}_{U=0} = 
  G(32') \eval{\fdv{(G^{-1})^\ee(33')}{U^\hh(12)}}_{U=0} G(3'1'),
\end{equation}
that stems from the differentiation of $\bG^{-1}$.

This set the stage for the introduction of the Gorkov-Dyson equation in the presence of the external pairing potential
\begin{multline}
  \label{eq:gorkov_dyson}
  \bG^{-1}(11') = \bG_0^{-1}(11') \\
  - \mqty(\Sigma^\he(11') & \Sigma^\hh(11') + U^\ee(11') \\ \Sigma^\ee(11') + U^\hh(11') & \Sigma^\eh(11')),
\end{multline}
which stems from the time derivative of $\bG$ (see Ref.~\onlinecite{Marie_2024b} for a detailed discussion).
This equation defines the four components of the self-energy in Nambu's formalism.
Using Eq.~\eqref{eq:gorkov_dyson}, the derivation can be pursued and one gets
\begin{equation}
  \label{eq:tmp_pp_bse}
\begin{split}
  K(12;1'2') 
  = & - G(32') \eval{\fdv{U^\hh(33')}{U^\hh(12)}}_{U=0} G(3'1') 
  \\
    & - G(32') \eval{\fdv{\Sigma^\ee(33')}{U^\hh(12)}}_{U=0} G(3'1'),
\end{split}
\end{equation}
where the first term in the right-hand side \titou{becomes the pp non-interacting propagator}
\begin{equation}
  \label{eq:K0}
  K_0(12;1'2') = \frac{1}{2} \qty[ G(11') G(22') - G(21') G(12') ].
\end{equation}
\titou{after the derivative is evaluated.}

\begin{figure}
  \centering
  \includegraphics[width=0.6\linewidth]{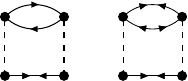}
  \caption{Diagrammatic representation of the two direct second-order terms contained in $\Sigma^\ee$.
  The dashed lines represent the Coulomb interaction, the solid lines with arrows denote the one-body propagator while the double-arrowed propagators represent $G^\hh$ and $G^\ee$.}
  \label{fig:second_order}
\end{figure}

Equation \eqref{eq:tmp_pp_bse} is transformed into a Dyson equation using the derivative chain rule.
This leads to four different terms because the chain rule has to be performed with respect to the four components of $\bG$.
However, only one of them is non-zero for a number-conserving system (\ie, at $U=0$) and this finally leads to the pp-BSE
\begin{multline}
  \label{eq:ppBSE}
  K(12;1'2') = K_0(12;1'2') \\
  - \int \dd(33'44') K(12;44') \Xi^\pp(44';33') K_0(33';1'2'),
\end{multline}
where the pp kernel,
\begin{equation}
  \label{eq:pp_kernel}
  \Xi^\pp(44';33') = \eval{\fdv{\Sigma^\ee(33')}{G^\ee(44')}}_{U=0},
\end{equation}
has been introduced.
The pp-BSE, defined in Eq.~\eqref{eq:ppBSE}, is represented diagrammatically in Fig.~\ref{fig:ppBSE}.
As one can readily seen from Eqs.~\eqref{eq:K0} and \eqref{eq:pp_kernel}, 
to compute $K$, one has to choose $G$ and the anomalous self-energy $\Sigma^\ee$ to build $K_0$ and $\Xi^\pp$, respectively.
This procedure is fully analog to the eh-BSE equation as discussed previously.
According to the definition of $\Xi^\pp$ in Eq.~\eqref{eq:pp_kernel}, the approximate expression of $\Sigma^\ee$ must contain exactly one anomalous propagator $G^\ee$ for the derivative to be non-zero.
To illustrate this, we examine the two direct second-order self-energy terms, with respect to the Coulomb interaction, contributing to the anomalous self-energy $\Sigma^\ee$. \cite{Soma_2011}
These diagrams are represented in Fig.~\ref{fig:second_order}.
The diagram on the right side possesses three anomalous propagators; thus, its derivative necessarily vanishes when evaluated at $U=0$.
In contrast, the first diagram has only one instance of $G^\ee$, resulting in a non-zero kernel.

\subsection{Eigenvalue problem}

With the general pp-BSE now derived, the next step is to reformulate it in a way suitable for implementation in standard quantum chemistry software. 
This involves first transforming the equations into frequency space, followed by projection into a spinorbital basis. 
Here, the indices $i,j,k,l,m$ denote occupied spinorbitals, $a,b,c,d,e$ are virtual spinorbitals, and $p,q,r,s$ corresponds to generic spinorbitals.
Finally, $\mu$ denotes a composite index $ia$ while $\nu$ corresponds to a composite index $ij$ (with $i<j$) or $ab$ (with $a<b$).

Before performing the Fourier transform, we recall that particle pairs are assumed to be created and annihilated instantaneously, \ie, $t_2 = t_{1}^+$ and $t_{2'} = t_{1'}^+$.
Hence, in Eq.~\eqref{eq:ppBSE}, $K(12;1'2')$ and $K_0(12;1'2')$ depend only on a single time difference, namely, $t_1 - t_{1'}$.
However, in the second term of the right-hand side of Eq.~\eqref{eq:ppBSE}, the quantities depend on two or three time differences.
These time dependencies are explicitly treated in the \SupInf. 
The Fourier transform leads to the following frequency-space pp-BSE
\begin{multline}
  \label{eq:FT_BSE}
  K(\omega) = K_0(\omega) 
  \\
  - \int\frac{\dd(\bar{\omega}\tilde{\omega})}{(2\pi)^2} K(\tilde{\omega},\omega) 
  \Xi^\pp(-\tilde{\omega},-\bar{\omega},\omega) K_0(\bar{\omega},\omega).
  \end{multline}
The two- and three-frequency Fourier transforms are defined in the \SupInf.
As can be readily seen, Eq.~\eqref{eq:FT_BSE} cannot be inverted directly.

To circumvent this issue, we rely on the procedure of Ref.~\onlinecite{Sangalli_2011} developed for the eh-BSE.
The pp-BSE is recast in an invertible form as
\begin{equation}
  \label{eq:invertible_ppBSE}
  K(\omega) = K_0(\omega) - K(\omega) \tilde{\Xi}^\pp(\omega) K_0(\omega),
\end{equation}
where the frequency-dependent kernel is
\begin{multline}
  \label{eq:dynamic_kernel}
  \tilde{\Xi}^\pp(\omega) = \int \frac{\dd(\tilde{\omega}\bar{\omega})}{(2\pi)^2} (K^{-1})(\omega) K(\bar{\omega},-\eta,\omega) 
  \\
  \times \Xi^\pp(-\tilde{\omega},-\bar{\omega},\omega) K_0(-\eta,\tilde{\omega},\omega) (K_0^{-1})(\omega).
\end{multline}
However, at this stage, the newly derived form is not yet practical as the kernel depends on $K$ and must, therefore, be solved iteratively.
Equation \eqref{eq:dynamic_kernel} is thus linearized by substituting $K$ by $K_0$.
\titou{This approximation might seem drastic but it has proven successful in the eh-BSE case. \cite{Sangalli_2011,Rebolini_2013,Loos_2020h}}
It is worth mentioning that if the initial kernel is static, \ie, $\Xi^\pp(\tilde{\omega},\bar{\omega},\omega) = \Xi^\pp$, the approximate dynamic kernel remains unchanged, \ie, $\tilde{\Xi}^\pp(\omega) = \Xi^\pp$.

We now express the pp-BSE as an eigenvalue problem.
Once inverted and projected in a finite basis set, Eq.~\eqref{eq:invertible_ppBSE} becomes
\begin{equation}
  K^{-1}_{pq,rs}(\omega) = (K_0^{-1})_{pq,rs}(\omega) + \tilde{\Xi}^\pp_{pq,rs}(\omega),
\end{equation}
where the four-index tensors have been transformed into matrices by defining composite indices.
Finally, one can show that finding the zeros of $K^{-1}(\omega)$, that is, the DIPs and DEAs, is equivalent to solving the following non-Hermitian non-linear eigenvalue problem
\begin{equation}
  \label{eq:pp_bse_eigenvalueeq}
  \mqty(\bC(\Omega_{n}) & \bB \\ - \bB^\dagger & - \bD(-\Omega_{n})) \cdot \mqty  ( \boldsymbol{X}_{n} \\ \boldsymbol{Y}_{n} ) = \Omega_{n}  \mqty( \boldsymbol{X}_{n} \\ \boldsymbol{Y}_{n}),
\end{equation}
where
\begin{equation}
  \begin{split}
  \label{eq:pp_bse_block}
    C_{ab,cd}(\omega) &= (\epsilon_a + \epsilon_b)\delta_{ac}\delta_{bd} + \ii\tilde{\Xi}^\pp_{ab,cd}(\omega),
    \\
    B_{ab,ij} &= + \ii \tilde{\Xi}^\pp_{ab,ij},
    \\
    D_{ij,kl}(-\omega) &= -(\epsilon_i + \epsilon_j)\delta_{ik}\delta_{jl} + \ii\tilde{\Xi}^\pp_{ij,kl} (-\omega),
  \end{split}
\end{equation}
with the following restrictions on the indices: $i<j$, $k<l$, $a<b$, and $c<d$.
These three blocks will be referred to as the ee-ee, ee-hh, and hh-hh blocks, respectively.
(In addition, note that $\tilde{\Xi}^\pp_{ab,ij}$ is, in general, frequency-dependent. 
However, for every kernel considered in this work, this coupling block is found to be static.)
Therefore, if the kernel is static, $\tilde{\Xi}^\pp_{ij,kl} (-\omega) = \Xi^\pp_{ij,kl}$ and $\tilde{\Xi}^\pp_{ab,cd} (\omega) = \Xi^\pp_{ab,cd}$, finding the poles of $K$ reduces to a generalized linear eigenvalue problem.
  Note the close connection between the static pp-BSE eigenvalue problem and the pairing TDDFT equations derived by van Aggelen and co-workers. \cite{vanAggelen_2013,vanAggelen_2014}
This connection is fully analogous to the one between eh-BSE and TDDFT. \cite{Blase_2018,Blase_2020}

In both the dynamic and static cases, the problem can be simplified by neglecting the (static) coupling block $\boldsymbol{B}$.
This widespread approximation is referred to as the Tamm-Dancoff approximation (TDA).
In this case, one ends up solving two independent Hermitian non-linear eigenvalue problems for the ee and hh sectors
  \begin{subequations}
    \begin{align}
      \label{eq:C_ppTDA}
      + \bC(\Omega_{n}) \cdot \bX_{n} & = \Omega_{n} \bX_{n},
      \\
      \label{eq:D_hhTDA}
      - \bD(-\Omega_{n}) \cdot \bY_{n} & = \Omega_{n} \bY_{n}.
    \end{align}
  \end{subequations}
Finally, both the static and dynamic eigenvalue problems can be spin-adapted to split them into independent problems of smaller size.
The spin adaptation is performed and discussed in detail in the \SupInf.

\subsection{Dynamical perturbative correction}

As mentioned earlier, if the kernel is dynamic, then finding the poles of $K$ requires solving a generalized non-linear eigenvalue problem [see Eq.~\eqref{eq:pp_bse_eigenvalueeq}].
Alternatively, the dynamical effects can be perturbatively accounted for in order to avoid the cumbersome non-linear procedure.
This is the choice that has been made in this work.
In addition, the dynamic perturbation is only considered within the TDA.
This dynamic perturbation has already been presented in detail in Ref.~\onlinecite{Loos_2020h} (see also Refs.~\onlinecite{Rohlfing_2000,Rebolini_2016}) for the eh-BSE case and shall only be outlined here for the hh sector.
An analogous procedure exists for the ee sector.

The matrix to diagonalize is decomposed into a static and dynamic part $\bD(\omega) = \bD^{(0)} + \bD^{(1)}(\omega)$.
Then, the static problem is first solved
\begin{equation}
  \bD^{(0)} \boldsymbol{Y}^{(0)}_{n} = - \Omega^{(0)}_{n} \boldsymbol{Y}^{(0)}_{n},
\end{equation}
and the perturbative correction to the $n$th eigenvalue $\Omega^{(0)}_{n}$ is computed as
\begin{equation}
  \Omega^{(1)}_{n} = (\boldsymbol{Y}^{(0)}_{n})^\dag \cdot \bD^{(1)}(-\Omega^{(0)}_{n}) \cdot \boldsymbol{Y}^{(0)}_{n}.
\end{equation}
Finally, the corrected eigenvalue is given by
\begin{equation}
  \Omega_{n} = \Omega^{(0)}_{n} + Z_{n} \Omega^{(1)}_{n},
\end{equation}
where the renormalization factor is
\begin{equation}
\label{eq:Z}
  Z_{n} = \left[1 - (\boldsymbol{Y}^{(0)}_{n})^\dag \cdot \eval{\pdv{\bD^{(1)}(-\omega)}{\omega}}_{\omega = \Omega^{(0)}_{n}} \cdot \boldsymbol{Y}^{(0)}_{n} \right]^{-1}.
\end{equation}

\section{Approximate kernels}
\label{sec:kernels}

Now that the general formalism has been discussed in detail, some practical approximations will be presented.
In particular, this section focuses on kernel approximations.
As mentioned in Sec.~\ref{sec:bse}, this is not the only approximation made in practice, as one must also choose an approximate form for the one-body propagator $G$.
The effect of both sources of approximations will be gauged in Sec.~\ref{sec:results}.

The present section is divided into various subsections, each corresponding to a given anomalous self-energy and its associated kernel.
First, we will show that considering the static Bogoliubov self-energy leads to the ubiquitous pp-RPA kernel.
Then, three kernels that go beyond pp-RPA will be presented.
These are the direct analogs of well-known kernels that have been considered for the eh-BSE: a kernel based on a self-energy correct up to second order in the Coulomb interaction and kernels based on the $GW$ and $T$-matrix self-energies.
The corresponding spin-adapted matrix elements for each of these kernels are reported in the \SupInf.

\subsection{The first-order Coulomb kernel}
\label{sec:1st_kernel}

The perturbation expansion of $\Sigma^\ee$ with respect to the Coulomb interaction has only one first-order term, which reads \cite{Soma_2011}
\begin{equation}
  \label{eq:bogol_self_nrj}
  \Sigma_{\text{B}}^\ee(11') = -\ii\int\dd{(33')} v(33'^{+};11'^{++}) G^{\ee}(33'),
\end{equation}
where
\begin{equation}
  v(11';22') = \delta(12)\frac{\delta(t_1 - t_{1'})}{\abs{\br_1-\br_{1'}}}\delta(1'2').
\end{equation}
The resulting kernel $\ii \Xi_{\text{B}}^\pp(11';22') = \bar{v}(11';22')/2 = [v(11';22') - v(11';2'2)]/2$ is simply the anti-symmetric Coulomb interaction.
Hence, this is a static kernel that can be used without any further approximation.
Once projected in a basis set, its matrix elements are simply $\ii (\Xi_{\text{B}}^\pp)_{pq}^{rs} = \mel{pq}{}{rs} = \braket{pq}{rs} - \braket{pq}{sr}$, where the two-electron integrals in the spin-orbital basis are defined as
\begin{equation}
  \braket{pq}{rs} = \iint \dd\bx_1 \dd\bx_2 \frac{\SO{p}(\bx_1) \SO{q}(\bx_2)\SO{r}(\bx_1) \SO{s}(\bx_2)}{\abs{\br_1 - \br_2}}.
\end{equation}
Note that here and throughout the manuscript, we assume real orbitals.
The corresponding matrix elements of the linear eigenproblem read
\begin{equation}
  \begin{split}
  \label{eq:pp_rpa}
    C_{ab,cd}^{\RPA} &= (\epsilon_a + \epsilon_b)\delta_{ac}\delta_{bd} + \mel{ab}{}{cd},
    \\
    B_{ab,ij}^{\RPA} &= \mel{ab}{}{ij},
    \\
    D_{ij,kl}^{\RPA} &= -(\epsilon_i + \epsilon_j)\delta_{ik}\delta_{jl} + \mel{ij}{}{kl},
  \end{split}
\end{equation}
which is easily recognized to be the well-known pp-RPA eigenvalue problem.

In the absence of instabilities, which corresponds in this case to test the stability of the reference $N$-electron state towards a $(N\pm2)$-electron state, \cite{Peng_2013,Scuseria_2013} the pp-RPA problem yields two sets of eigenvalues: a set of positive eigenvalues corresponding to DEAs [$(N+2)$-electron states], and a set of negative eigenvalues corresponding to DIPs [$(N-2)$-electron states].

\subsection{The second-order Coulomb kernel}
\label{sec:2nd_kernel}

As explained in Sec.~\ref{sec:bse}, there are two second-order direct terms in the perturbative expansion of $\Sigma^\ee$ and only one contributes to $\Xi^\pp$.
This self-energy diagram is represented on the left side of Fig.~\ref{fig:second_order}.
$\Sigma^\ee$ also contains two exchange terms and, for the same reason, only one yields a non-zero kernel.
Once added up with the first-order term of Eq.~\eqref{eq:bogol_self_nrj}, the second-order kernel reads \cite{Soma_2011}
\begin{equation}
  \label{eq:2nd_self_nrj}
  \Sigma^{\ee,(2)}(11') = -\ii W^{(2)}(33';11')G^\ee(33'),
\end{equation}
where the second-order effective interaction,
\begin{multline}
  W^{(2)}(33';11') = v(33';11') 
  \\
  - \ii\bar{v}(34';14)L_0(42;4'2')\bar{v}(2'3';21'),
\end{multline}
has been introduced.
It corresponds to an antisymmetrized Coulomb interaction ``screened'' up to second order in the Coulomb interaction.
The differentiation of $\Sigma^{\ee,(2)}$ with respect to $G^\ee$ yields
\begin{equation}
  \label{eq:gf2_kernel}
  \ii \Xi^{\pp,(2)}(11';22') = \frac{1}{2} [W^{(2)}(11';22') - W^{(2)}(11';2'2)].
\end{equation}

Because this kernel is frequency-dependent, one must either evaluate it at $\omega = 0$ or the corresponding approximate effective dynamic kernel has to be computed [see Eq.~\eqref{eq:dynamic_kernel}].
The first choice leads to
\begin{subequations}
  \label{eq:kernel_elements_gf2}
  \begin{align}
    \ii \Xi^{\pp,(2)}_{ab,cd} &= W^{(2)}_{abcd} - W^{(2)}_{abdc}, 
    \\
    \ii \Xi^{\pp,(2)}_{ab,ij} &= W^{(2)}_{abij} - W^{(2)}_{abji}, 
    \\
    \ii \Xi^{\pp,(2)}_{ij,kl} &= W^{(2)}_{ijkl} - W^{(2)}_{ijlk},
  \end{align}
\end{subequations}
where $W^{(2)}_{pqrs}$ is a short notation for $W^{(2)}_{pqrs}(\omega=0)$.
The matrix elements of the frequency-dependent effective interaction are given by
\begin{multline}
  W^{(2)}_{pqrs}(\omega) = \braket{pq}{rs} + 
  \\
  \sum_{me} \qty[ \frac{\mel{pm}{}{re} \mel{eq}{}{ms}}{\omega - (\epsilon_e - \epsilon_m - \ii\eta)} - \frac{\mel{pe}{}{rm} \mel{mq}{}{es}}{\omega - (\epsilon_m - \epsilon_e + \ii\eta)} ].
\end{multline}
The dynamic kernel $\tilde{\Xi}^{\pp,(2)}$ is discussed in the \SupInf.

\subsection{The $GW$ kernel}
\label{sec:gw_kernel}

In the case of the eh propagator, the $GW$ kernel is arguably the most successful in the context of quantum chemistry and condensed matter physics.
Hence, a $GW$-like kernel for the pp channel is a natural target but it does require a $GW$-like self-energy approximation for $\Sigma^\ee$.
Such quantity can be obtained by generalizing Hedin's equations to the Gorkov propagator, as shown in Ref.~\onlinecite{PhDEssenberger}.
For the sake of completeness, the derivation of $\Sigma^{\ee,GW}$ is also reported in the \SupInf.

The resulting self-energy expression is
\begin{equation}
  \label{eq:gw_self_nrj}
  \Sigma^{\ee,GW}(11') = -\ii G^\ee(33') W(33';11'),
\end{equation}
and the associated kernel is 
\begin{equation}
  \label{eq:gw_kernel}
  \ii \Xi^{\ee,GW}(11';22') = \frac{1}{2} [W(11';22') - W(11;2'2)],
\end{equation}
where $W$ is the screened interaction computed at the eh-RPA level 
\begin{multline}
  W(11';22') = v(11';22') \\
  - \ii W(13;23') L_0(4'3';43) v(1'4;2'4').
\end{multline}
Note that, for notational convenience, we use the same notation for $W$ in Eq.~\eqref{eq:gw_self_nrj} and for $W$ in Eq.~\eqref{eq:gw_kernel}.
However, the first contains both normal and anomalous bubbles and the latter only includes the normal bubbles as one sets $U\to0$ according to Eq.~\eqref{eq:pp_kernel}.

We also note that contrary to the eh $GW$ kernel where one usually neglects the term $\fdv*{W}{G}$ (which can be shown to be of second-order in $W$), \cite{Baym_1961,Yamada_2022,Monino_2023} this derivative is effectively zero for the pp channel as the anomalous propagator vanishes at $U=0$.
Another interesting property of this kernel is that it justifies the \textit{ad hoc} kernel of Rohlfing and co-workers. \cite{Deilmann_2016}
Indeed, while in the eh case only the exchange term is screened, the authors of Ref.~\onlinecite{Deilmann_2016} argued that one has to screen both the Hartree and exchange in the pp kernel using symmetry arguments.
Therefore, Eq.~\eqref{eq:gw_kernel} provides a first-principle justification for this choice.

In the static approximation, the matrix elements of the $GW$ kernel are
\begin{subequations}
  \label{eq:pp_bse_gw}
  \begin{align}
    \ii \Xi^{\pp,GW}_{ab,cd} &= W_{abcd} - W_{abdc},
    \\
    \ii \Xi^{\pp,GW}_{ab,ij} &= W_{abij} - W_{abji}, 
    \\
    \ii \Xi^{\pp,GW}_{ij,kl} &= W_{ijkl} - W_{ijlk},
  \end{align}
\end{subequations}
with $W_{pqrs} = W_{pqrs}(\omega=0)$. 
The elements of the dynamically-screened interaction are given by
\begin{equation}
  W_{pqrs}(\omega) = \braket{pq}{rs} + \sum_\mu \qty[\frac{M_{pr,\mu} M_{sq,\mu}}{\omega - \Omega_\mu + \ii \eta} - \frac{M_{rp,\mu} M_{qs,\mu}}{\omega + \Omega_\mu - \ii \eta}],
\end{equation}
where the transition densities are
\begin{equation}
  M_{pq,\mu} =\sum_{ia} \qty[ X_{ia,\mu} \braket{ap}{iq} + Y_{ia,\mu} \braket{ip}{aq} ].
\end{equation}
Here, $X_{pq,\mu}$, $Y_{pq,\mu}$, and $\Omega_\mu$ are the matrix elements of the eigenvectors and the eigenvalues of the (direct) eh-RPA problem, which is reported in the \SupInf.
\titou{Hence, the static ppBSE@$GW$ eigenvalue problem is the same as the ppRPA one with screened two-electron integrals instead of the usual unscreened ones.}

To conclude this subsection, the expression of the hh-hh $GW$ effective dynamic kernel matrix elements are reported
  \begin{equation}
    \begin{split}
      \label{eq:gw_dynamic_kernel}
      \ii \tilde{\Xi}^{\pp,GW}_{ij,kl}(-\omega) 
      & = \mel{ij}{}{kl} 
      \\
      & + \sum_\mu \frac{M_{ik,\mu} M_{lj,\mu} - M_{jk,\mu} M_{li,\mu}}{\omega - (-\epsilon_j - \epsilon_l + \Omega_\mu - \ii\eta)} 
      \\
      & + \sum_\mu \frac{M_{ki,\mu} M_{jl,\mu} - M_{kj,\mu} M_{il,\mu}}{\omega - (-\epsilon_i - \epsilon_k + \Omega_\mu - \ii\eta)}  
      \\
      & + \sum_\mu \frac{M_{ik,\mu} M_{lj,\mu} - M_{jk,\mu} M_{li,\mu}}{\omega - (-\epsilon_i - \epsilon_l + \Omega_\mu - \ii\eta)} 
      \\
      & + \sum_\mu \frac{M_{ki,\mu} M_{jl,\mu} - M_{kj,\mu} M_{il,\mu}}{\omega - (-\epsilon_j - \epsilon_k + \Omega_\mu - \ii\eta)} 
    \end{split}
  \end{equation}
The ee-hh and ee-ee blocks are reported in the \SupInf.

\subsection{The $T$-matrix kernel}
\label{sec:gt_kernel}

Finally, the kernel based on the pp $T$-matrix effective interaction is discussed.
The $T$-matrix anomalous self-energy approximation has been derived in a previous study by some of the authors \cite{Marie_2024b}
(see also Ref.~\onlinecite{Bozek_2002}).
However, the associated pp kernel vanishes in the normal phase.
Hence, there is no kernel of first order in $T$ for the pp-BSE.
This could have been anticipated as the pp $T$-matrix is already based on pp-RPA. \cite{Marie_2024b}

Computing the first vertex correction to this self-energy leads to an anomalous self-energy of second order in $T$ which contains exactly one anomalous propagator \cite{Marie_2024b}
\begin{equation}
  \label{eq:g3t2_self_nrj}
  \Sigma^{\ee,(2T)}(11') = -\ii W^{(2T)}(33';11')G^\ee(33'),
\end{equation}
where the second-order effective interaction,
\begin{multline}
  W^{(2T)}(33';11') = v(33';11') 
  \\
  - \ii T(34';14) L_0(42;4'2') T(2'3';21'),
\end{multline}
Note that this self-energy is equivalent to the second-order self-energy of Eq.~\eqref{eq:2nd_self_nrj}, where the antisymmetrized Coulomb interaction is replaced by the effective interaction $T$ and thus leads to a kernel that is non-zero in the normal phase
\begin{equation}
  \label{eq:gf2_kernel}
  \ii \Xi^{\pp,(2T)}(22';11') = \frac{1}{2} [W^{(2T)}(22';11') - W^{(2T)}(2'2;11')],
\end{equation}
where
\begin{multline}
  W^{(2T)}(33';11') = v(33';11') \\
  -\ii T(34';14)L_0(42;4'2')T(2'3';21')
\end{multline}
is an effective interaction analog to $W^{(2)}$, but of second order in $T$ instead of $v$.
In the following, this kernel is computed under the approximation of a static effective interaction $T$. 
The derivation of the second-order kernel in a frequency-dependent effective interaction is beyond the scope of this work.

The matrix elements of the kernel $\Xi^{\pp,(2T)}$ are the same as in Eq.~\eqref{eq:kernel_elements_gf2} but with $W^{(2T)}$ instead of $W^{(2)}$.
The matrix elements of this effective interaction are given by
\begin{multline}
  \label{eq:w_2t}
  W^{(2T)}_{pqrs}(\omega) = \braket{pq}{rs} \\
  +  \sum_{me} \qty[ \frac{T_{pmre} T_{eqms}}{\omega - (\epsilon_{e} - \epsilon_{m} - \ii\eta)} - \frac{T_{perm} T_{mqes}}{\omega - (\epsilon_{m} - \epsilon_{e} + \ii\eta)} ],
\end{multline}
where $T_{pqrs} = T_{pqrs}(\omega = 0)$ and the pp $T$-matrix elements are
\begin{multline}
  \label{eq:tensor_elem_t}
  T_{pqrs}(\omega) = \mel{pq}{}{rs} \\
  + \sum_{\nu} \frac{M_{pq,\nu}^{N+2} M_{rs,\nu}^{N+2}}{\omega - \Omega_\nu^{N+2} + \ii\eta} - \sum_{\nu} \frac{M_{pq,\nu}^{N-2} M_{rs,\nu}^{N-2}}{\omega - \Omega_\nu^{N-2} - \ii\eta}.
\end{multline}
Finally, the transition densities are defined as
\begin{subequations}
  \begin{align}
    M_{pq,\nu}^{N+2} &= \sum_{c<d} \mel{pq}{}{cd} X_{cd,\nu}^{N+2} + \sum_{k<l} \mel{pq}{}{kl} Y_{kl,\nu}^{N+2},
    \\
    M_{pq,\nu}^{N-2} &= \sum_{c<d} \mel{pq}{}{cd} X_{cd,\nu}^{N-2} + \sum_{k<l} \mel{pq}{}{kl} Y_{kl,\nu}^{N-2},
  \end{align}
\end{subequations}
where $X^{N\pm 2}_{pq,\nu}$, $Y^{N\pm 2}_{pq,\nu}$, and $\Omega_\nu^{N\pm 2}$ are the matrix elements of the eigenvectors and the eigenvalues of the pp-RPA problem and are explicitly defined in the \SupInf.

\section{Computational details}
\label{sec:comp_det}

The set of molecules considered here is the same as in the recent benchmark of valence IPs and satellite transitions developed by some of the authors. \cite{Marie_2024}
This work extends this benchmark by computing the exact lowest DIPs (singlet and triplet) of these 23 molecules.
The calculations are performed in Dunning's aug-cc-pVTZ. \cite{Dunning_1989,Kendall_1992,Prascher_2011,Woon_1993} 
The geometries are extracted from the \textsc{quest} database, \cite{Loos_2020d,Veril_2021} meaning that they have been optimized at the CC3 level \cite{Christiansen_1995b,Koch_1997} in the aug-cc-pVTZ basis set without frozen-core approximation.
These reference values are computed using the \textsc{quantum package} implementation \cite{Garniron_2019} of the configuration interaction using a perturbative selection made iteratively (CIPSI) method.\cite{Huron_1973,Giner_2013,Giner_2015,Caffarel_2016b,Garniron_2017,Garniron_2018}
The lowest DIPs of each molecule are obtained as energy differences between the neutral ground-state energy and the singlet and triplet cation ground-state energies.
\titou{The frozen-core approximation has been used for both calculations. Note that the $1s$ orbital of \ce{Li} and \ce{Be} was not frozen.}
We refer the reader to Ref.~\onlinecite{Marie_2024} for a detailed discussion of the CIPSI extrapolation procedure which is employed to produce the full configuration interaction (FCI) estimates.

Additionally, the single-site double core ionizations of four molecules (\ce{H2O}, \ce{NH3}, \ce{CH4}, and \ce{CO}) have been considered.
In a single-site double core hole (DCH) state, the two $1s$ electrons are ionized from the same orbital, while in a double-site DCH state, the ionization process occurs on distinct $1s$ orbitals.
Here, we only consider single-site DCHs.
These calculations are performed in Dunning's aug-cc-pCVTZ. \cite{Dunning_1989,Kendall_1992,Woon_1995}
The reference values have been computed under the core-valence separation (CVS) approximation, which restricts the CI expansion \titou{of the cation} to determinants with two core holes. \cite{Cederbaum_1980a,Ferte_2023}
\titou{In this case, the frozen-core approximation was not enforced for the neutral ground-state calculations.}

The various flavors of pp-BSE considered in Sec.~\ref{sec:results} have been implemented in an open-source in-house program, named \textsc{quack}. \cite{QuAcK}
The implementation relies on a full diagonalization of the spin-adapted pp-BSE matrices.
The machinery developed to reduce the cost of pp-RPA, such as Davidson diagonalization \cite{Yang_2014a} or active spaces, \cite{Zhang_2016,Li_2023} could be transposed to pp-BSE but this is beyond the scope of this work.
In addition, for some starting points and kernels, the TDA had to be enforced for \ce{BN} and \ce{C2} as the pp-BSE eigenproblem is unstable, that is, the number of negative eigenvalues is larger than the number of DIPs. \cite{Scuseria_2013}
Hence, for the sake of consistency, each \ce{BN} and \ce{C2} DIP has been computed within the TDA.
\titou{Note that this instability is expected in these systems as a consequence of the large negative energy of the lowest unoccupied molecular orbital (LUMO).}

The underlying one-body Green's function calculations [$GW$, GF(2), and $T$-matrix] calculations rely on the linearized version of the quasiparticle equation (without self-consistency) to obtain the quasiparticle energies. \cite{Marie_2024a}
The infinitesimal $\eta$ is set to zero for all calculations except for the perturbative correction, where a value of \SI{0.05}{\hartree} has been used.
The starting point of the one-body Green's calculations is always a (restricted) Hartree-Fock (HF) solution.

The $\Delta$SCF calculations have been performed with \textsc{quantum package}.
The determinants with two core holes have been optimized using the maximum overlap method (MOM). \cite{Gilbert_2008,Barca_2014,Barca_2018a}
Finally, the DIP-EOM-CCSD calculations have been performed using \textsc{q-chem 6.2.1}, \cite{Epifanovsky_2021} \titou{ and the CVS-DIP-EOM-CCSD using \textsc{ccpy}. \cite{ccpy} }

\begin{squeezetable}
\begin{table*}
  \caption{DIPs (in \si{\eV}) toward the singlet (left panel) and triplet (right panel) dication ground states in the aug-cc-pVTZ basis set computed at the FCI level and the pp-RPA level using various one-body energies: HF, $GW$, $T$-matrix, and GF(2).}
  \label{tab:tab1}
  \begin{ruledtabular}
    \begin{tabular}{lcccccccccc}
               &                \mc{5}{c}{Singlet DIPs}                   &                \mc{5}{c}{Triplet DIPs}                   \\
                                \cline{2-6}                                                      \cline{7-11}
      Molecule &  FCI  & ppRPA@HF & ppRPA@$GW$ & ppRPA@$GT$ & ppRPA@GF(2) &  FCI  & ppRPA@HF & ppRPA@$GW$ & ppRPA@$GT$ & ppRPA@GF(2) \\
      \hline                                                                                                             
      \ce{H2O} & 41.43 &    47.00 &      45.01 &      43.94 &       42.39 & 40.29 &    46.18 &      44.37 &      43.27 &       41.80 \\
      \ce{HF}  & 50.69 &    57.85 &      54.93 &      53.90 &       51.87 & 47.90 &    55.62 &      52.71 &      51.67 &       49.65 \\
      \ce{Ne}  & 65.43 &    72.80 &      69.34 &      68.64 &       66.63 & 62.19 &    70.12 &      66.65 &      65.96 &       63.95 \\
      \ce{CH4} & 38.98 &    41.07 &      40.91 &      39.95 &       39.61 & 38.27 &    40.37 &      40.21 &      39.25 &       38.91 \\
      \ce{NH3} & 35.91 &    39.45 &      38.45 &      37.47 &       36.55 & 38.38 &    41.98 &      41.25 &      40.23 &       39.44 \\
      \ce{CO}  & 41.29 &    43.99 &      42.71 &      41.71 &       40.75 & 41.56 &    42.61 &      41.94 &      41.06 &       40.54 \\
      \ce{N2}  & 42.81 &    46.27 &      44.20 &      42.93 &       41.53 & 43.70 &    46.66 &      46.14 &      45.24 &       44.96 \\
      \ce{BF}  & 34.53 &    34.72 &      35.25 &      34.50 &       34.67 & 38.47 &    39.33 &      38.10 &      37.29 &       36.44 \\
      \ce{LiF} & 40.00 &    46.75 &      43.69 &      42.77 &       40.47 & 37.35 &    44.71 &      41.66 &      40.73 &       38.44 \\
      \ce{BeO} & 32.18 &    36.43 &      34.97 &      33.73 &       32.00 & 30.17 &    35.06 &      33.60 &      32.36 &       30.63 \\
      \ce{BN}  & 34.98 &    36.34 &      36.73 &      36.25 &       35.29 & 33.73 &    35.19 &      35.59 &      35.10 &       34.15 \\
      \ce{C2}  & 35.98 &    37.58 &      38.58 &      38.74 &       38.88 & 35.14 &    36.50 &      37.50 &      37.66 &       37.80 \\
      \ce{CS}  & 33.36 &    34.87 &      34.76 &      33.80 &       33.71 & 32.68 &    34.22 &      34.11 &      33.15 &       33.06 \\
      \ce{LiCl}& 30.83 &    33.18 &      32.48 &      32.03 &       31.62 & 29.35 &    31.80 &      31.10 &      30.65 &       30.24 \\
      \ce{F2}  & 44.48 &    48.90 &      45.03 &      43.15 &       40.63 & 43.87 &    48.63 &      44.79 &      42.91 &       40.41 \\
      \ce{H2S} & 31.75 &    32.81 &      32.82 &      32.19 &       32.19 & 32.77 &    34.16 &      34.12 &      33.45 &       33.38 \\
      \ce{PH3} & 31.15 &    31.94 &      32.25 &      31.47 &       31.57 & 32.38 &    33.22 &      33.33 &      32.58 &       32.54 \\
      \ce{HCl} & 37.17 &    39.17 &      38.75 &      38.21 &       38.03 & 35.61 &    37.70 &      37.28 &      36.74 &       36.56 \\
      \ce{Ar}  & 44.74 &    47.14 &      46.38 &      45.98 &       45.73 & 42.98 &    45.42 &      44.66 &      44.26 &       44.01 \\
      \ce{SiH4}& 32.78 &    33.74 &      33.71 &      33.07 &       32.94 & 32.65 &    33.52 &      33.49 &      32.85 &       32.72 \\
      \ce{CH2O}& 33.33 &    35.66 &      34.20 &      32.79 &       31.45 & 35.68 &    38.35 &      37.61 &      36.41 &       35.64 \\
      \ce{CO2} & 38.54 &    40.26 &      38.96 &      37.69 &       36.72 & 37.37 &    39.83 &      38.54 &      37.27 &       36.32 \\
      \ce{BH3} & 36.72 &    37.50 &      37.62 &      36.95 &       36.84 & 35.44 &    36.14 &      36.32 &      35.62 &       35.52 \\
      \hline                                                                                                           
      MSE      &       &   \z2.89 &     \z1.85 &     \z0.99 &      \z0.13 &       &   \z3.02 &     \z2.05 &     \z1.21 &      \z0.40 \\
      MAE      &       &   \z2.89 &     \z1.85 &     \z1.23 &      \z0.96 &       &   \z3.02 &     \z2.08 &     \z1.45 &      \z1.06 \\
      RMSE     &       &   \z3.55 &     \z2.16 &     \z1.58 &      \z1.32 &       &   \z3.76 &     \z2.45 &     \z1.83 &      \z1.35 \\
      SDE      &       &   \z2.11 &     \z1.13 &     \z1.25 &      \z1.34 &       &   \z2.30 &     \z1.37 &     \z1.41 &      \z1.32 \\
      Min      &       &   \z0.19 &     \z0.42 &      -1.32 &       -3.85 &       &   \z0.70 &      -0.37 &      -1.18 &       -3.45 \\
      Max      &       &   \z7.37 &     \z4.24 &     \z3.21 &      \z2.90 &       &   \z7.93 &     \z4.81 &     \z3.78 &      \z2.67 \\
    \end{tabular}
  \end{ruledtabular}
\end{table*}
\end{squeezetable}

\begin{figure*}
  \centering
  \includegraphics[width=0.245\linewidth]{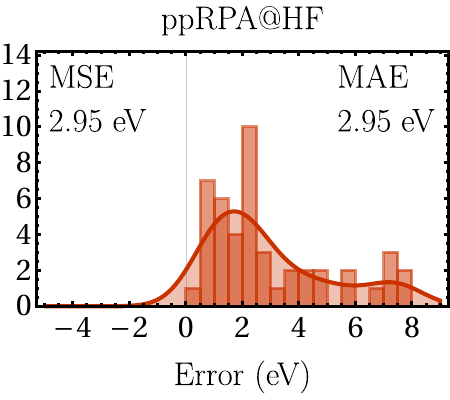}
  \includegraphics[width=0.245\linewidth]{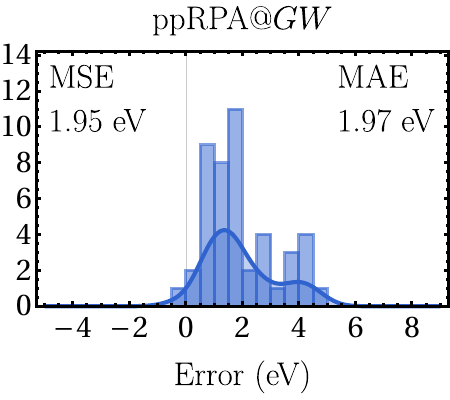}
  \includegraphics[width=0.245\linewidth]{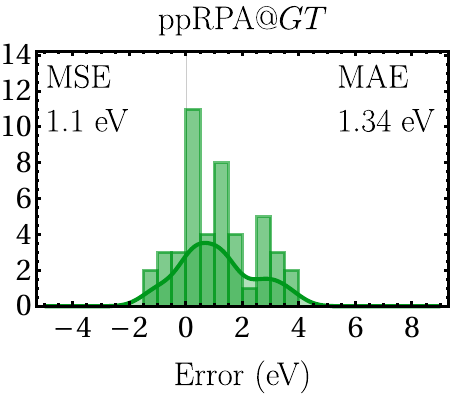}
  \includegraphics[width=0.245\linewidth]{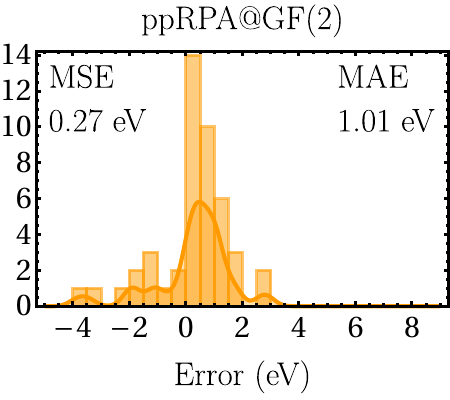}
  \caption{Histogram of the errors (with respect to FCI) for the singlet and triplet principal DIP of 23 small molecules computed in the aug-cc-pVTZ basis set at the pp-RPA level using various one-body energies: HF, $GW$, $T$-matrix, and GF(2).}
   \label{fig:error_plot_rpa}
\end{figure*}

\section{Results}
\label{sec:results}

Doubly ionized molecules can be obtained through various processes such as direct single- or two-photon ionizations or indirect mechanisms like the Auger-Meitner effect. \cite{Meitner_1922,Auger_1923}
The theoretical understanding of the dication electronic structure is a crucial tool to interpret the corresponding experimental spectra precisely.
For example, in a series of papers, Schirmer, Cederbaum, and co-workers applied the algebraic diagrammatic construction (ADC) \cite{Dreuw_2015,Schirmer_2018} to the pp propagator to compute valence DIPs. \cite{Schirmer_1984,Tarantelli_1989}
The pp-ADC became, after the seminal $\Delta$SCF study of Ortenburger and Bagus, \cite{Ortenburger_1975} the method of choice to decipher experimental Auger spectra. \cite{Tarantelli_1985,Tarantelli_1985a,Tarantelli_1985b,Tarantelli_1987,Tarantelli_1991,Pernpointner_2012}
Various other pp-propagator-based methods have also been considered to study this problem. \cite{Graham_1991,Liegener_1996,Noguchi_2005,Noguchi_2007,Noguchi_2008,Ida_2008}
Recently, highly correlated methods in combination with non-Hermitian extensions of quantum chemistry have been employed to go beyond the pp-ADC state-of-the-art for Auger spectra. \cite{Matz_2022,Matz_2023,Jayadev_2023,Ferino-Perez_2024}
In a different context, a clear theoretical understanding of DCH states allowed to precisely interpret the satellites of the corresponding spectra, \cite{Ferte_2020,Ferte_2022} as well as understanding the dynamics of these DCH states. \cite{Marchenko_2018,Ismail_2024}

In the following section, the pp-BSE formalism is employed to compute valence and core DIPs.
The accuracy of its various variants is assessed and discussed for both types of states.

\subsection{Valence double ionization potentials}
\label{sec:dip}

The first testbed of this study, designed to assess the various approximations of pp-BSE, is composed of the lowest singlet and triplet DIP of 23 small molecules.
The set of molecules is the same as in the valence IP and satellite benchmark reported in Ref.~\onlinecite{Marie_2024}.
The FCI-quality reference DIPs have been obtained using the CIPSI algorithm and are reported in Table \ref{tab:tab1}.
\titou{In addition, the main configuration of each CI expansion is reported in the \SupInf.}
As mentioned earlier, there are two levels of approximations in pp-BSE, namely, the underlying one-body orbital energies and the choice of the kernel.

The impact of the one-body propagators is investigated first.
The DIPs are computed with the pp-RPA first-order Coulomb kernel for various starting points, hence different orbital energies: HF, $GW$, $T$-matrix, and GF(2).
These are referred to as ppRPA@[HF, $GW$, $GT$, and GF(2)] and their respective values are gathered in Table \ref{tab:tab1}.
Figure \ref{fig:error_plot_rpa} presents the corresponding histograms of errors with respect to FCI.
Before discussing these errors for DIPs, one should recall the error on single IPs of these four starting points.
For a set of 58 valence IPs (of these 23 molecules), the mean absolute errors (MAEs) of these four methods are $1.30$, $0.47$, $0.49$, and \SI{0.81}{\eV} while the mean signed errors (MSEs) are $1.23$, $0.40$, $0.01$, and \SI{-0.55}{\eV}. \cite{Marie_2024}

The MAE and MSE associated with the pp-RPA@HF DIPs are \titou{both \SI{2.95}{\eV}.}
This error is larger than the \SI{1.30}{\eV} MAE of HF IPs.
However, the mean value of IPs and DIPs on these two sets are $16.29$ and \SI{38.41}{\eV}, respectively.
Hence, the ratios of MAE over mean value are similar: $0.07$ for IPs and $0.08$ for DIPs.

The $GW$ one-body energies offer a quantitative improvement by decreasing the MAE by approximately \SI{1}{\eV}.
In addition, the positive MSE, \SI{1.95}{\eV}, almost equal to the MAE, \SI{1.97}{\eV}, is is in agreement with the ppRPA@$GW$ results of Noguchi and co-workers who also observed a systematic overestimation of DIPs for several small molecules. \cite{Noguchi_2005,Noguchi_2007,Noguchi_2008}
The ppRPA@$GT$ and ppRPA@GF(2) have a similar MAE, $1.34$ and \SI{1.01}{\eV}, respectively.
However, with a MSE of \SI{0.27}{\eV}, the ppRPA@GF(2) error distribution is more centered around zero than its ppRPA@$GT$ counterpart with its MSE of \SI{1.10}{\eV}.
Note that the decrease of the MSE along the series HF, $GW$, $T$-matrix, and GF(2) is the same as observed in the case of single IPs.
The error spread, quantified by the root mean square error (RMSE) and the standard deviation error (SDE), is also decreased when using $GW$, $T$-matrix, and GF(2) orbital energies rather than HF.
Finally, for these four methods, there is no noticeable difference between the MAE for singlets and triplets (see Table \ref{tab:tab1}).

Note that, for the $GW$, GF(2), and $T$-matrix self-energies, we rely on the one-shot scheme to compute the quasiparticle energies.
The effect of self-consistency on the one-body energies has also been gauged by computing the pp-RPA DIPs based on ev$GW$ \cite{Shishkin_2007a,Blase_2011b,Marom_2012,Wilhelm_2016,Kaplan_2016} and qs$GW$ \cite{Kaplan_2016,Faleev_2004,vanSchilfgaarde_2006,Kotani_2007,Ke_2011,Marie_2023} starting points.
The corresponding values are reported in the \SupInf and evidence that self-consistency slightly improves both the IPs and DIPs.

Now that the effect of the one-body energies has been discussed in detail, approximate kernels going beyond the first-order static approximation will be considered.
These kernels will be compared with the state-of-the-art wave function method for valence DIP, namely the DIP-EOM-CCSD.
The corresponding values are reported in Table \ref{tab:tab2}.
For this benchmark set, DIP-EOM-CCSD has a MAE of \SI{0.61}{\eV} and a MSE of \SI{0.61}{\eV}.
We recall that its computational cost is $\order*{N^6}$ where $N$ is the size of the one-body basis set.

\begin{squeezetable}
\begin{table*}
  \caption{DIPs (in \si{\eV}) toward the singlet (left panel) and triplet (right panel) dication ground states in the aug-cc-pVTZ basis set for DIP-EOM-CCSD and three variants of pp-BSE corresponding to a static $GW$ kernel (ppBSE@$GW$), a static $GW$ kernel within the TDA (TDA@ppBSE@$GW$) and a dynamic $GW$ kernel within the TDA (TDA@dynBSE@$GW$).
  The renormalization factor associated with the dynamic correction, as defined in Eq.~\eqref{eq:Z}, is reported in parentheses.}
  \label{tab:tab2}
  \begin{ruledtabular}
    \begin{tabular}{lcccccccc}
               &              \mc{4}{c}{Singlet DIPs}             &              \mc{4}{c}{Triplet DIPs}             \\
               \cline{2-5}                                        \cline{6-9}
               &          &            &     TDA    &     TDA     &          &            &     TDA    &     TDA     \\
      Molecule & DIP-CCSD & ppBSE@$GW$ & ppBSE@$GW$ & dynBSE@$GW$ & DIP-CCSD & ppBSE@$GW$ & ppBSE@$GW$ & dynBSE@$GW$ \\
      \hline
      \ce{H2O} &    42.04 &      40.30 &      40.48 & 41.04(0.78) &    41.06 &      40.25 &      40.30 & 41.17(0.85) \\
      \ce{HF}  &    51.53 &      49.22 &      49.35 & 50.22(0.80) &    48.89 &      47.27 &      47.33 & 48.52(0.85) \\
      \ce{Ne}  &    66.16 &      63.15 &      63.30 & 64.43(0.84) &    63.03 &      60.79 &      60.86 & 62.28(0.88) \\
      \ce{CH4} &    39.29 &      39.31 &      39.37 & 39.78(0.94) &    38.59 &      38.76 &      38.79 & 39.18(0.94) \\
      \ce{NH3} &    36.25 &      34.94 &      35.13 & 35.53(0.81) &    38.86 &      38.61 &      38.65 & 39.38(0.92) \\
      \ce{CO}  &    42.02 &      41.83 &      41.98 & 42.13(0.90) &    42.19 &      41.57 &      41.59 & 41.68(0.97) \\
      \ce{N2}  &    43.94 &      44.10 &      44.16 & 44.19(0.96) &    44.57 &      44.34 &      44.37 & 44.86(0.94) \\
      \ce{BF}  &    34.86 &      33.54 &      33.77 & 33.68(0.81) &    38.59 &      38.01 &      38.02 & 38.14(0.98) \\
      \ce{LiF} &    41.05 &      37.75 &      37.88 & 38.63(0.88) &    38.55 &      36.00 &      36.05 & 36.83(0.82) \\
      \ce{BeO} &    33.21 &      30.56 &      30.67 & 31.02(0.81) &    31.38 &      29.39 &      29.43 & 29.94(0.78) \\
      \ce{BN}  &    35.33 &      33.94 &      33.94 & 34.27(0.83) &    34.17 &      32.98 &      32.98 & 33.41(0.86) \\
      \ce{C2}  &    37.18 &      36.44 &      36.44 & 36.77(0.88) &    36.34 &      35.52 &      35.52 & 35.91(0.90) \\
      \ce{CS}  &    34.02 &      33.79 &      33.84 & 34.09(0.95) &    33.32 &      33.33 &      33.35 & 33.56(0.96) \\
      \ce{LiCl}&    31.36 &      30.09 &      30.17 & 30.41(0.88) &    29.93 &      29.02 &      29.05 & 29.34(0.86) \\
      \ce{F2}  &    45.41 &      45.05 &      45.08 & 45.19(0.99) &    44.85 &      44.55 &      44.57 & 44.85(0.98) \\
      \ce{H2S} &    31.99 &      31.03 &      31.18 & 31.44(0.88) &    33.09 &      32.79 &      32.82 & 33.06(0.92) \\
      \ce{PH3} &    31.37 &      30.91 &      31.05 & 31.27(0.90) &    32.61 &      32.62 &      32.64 & 32.78(0.95) \\
      \ce{HCl} &    37.52 &      36.57 &      36.66 & 37.06(0.89) &    36.01 &      35.41 &      35.44 & 35.87(0.91) \\
      \ce{Ar}  &    45.17 &      44.01 &      44.10 & 44.57(0.90) &    43.45 &      42.66 &      42.70 & 43.18(0.92) \\
      \ce{SiH4}&    33.01 &      33.36 &      33.38 & 33.49(0.98) &    32.90 &      33.20 &      33.22 & 33.30(0.98) \\
      \ce{CH2O}&    33.85 &      33.15 &      33.24 & 33.52(0.94) &    36.53 &      35.84 &      35.86 & 36.41(0.92) \\
      \ce{CO2} &    39.10 &      38.45 &      38.50 & 38.79(0.97) &    38.16 &      37.87 &      37.90 & 38.27(0.97) \\
      \ce{BH3} &    36.90 &      36.72 &      36.78 & 36.97(0.95) &    35.63 &      35.61 &      35.64 & 35.81(0.96) \\
      \hline                                                                                                
      MSE      &   \z0.59 &      -0.47 &      -0.38 &    -0.03\zs &   \z0.64 &      -0.07 &      -0.04 &   \z0.33\zs \\
      MAE      &   \z0.59 &     \z0.84 &     \z0.78 &   \z0.61\zs &   \z0.64 &     \z0.48 &     \z0.48 &   \z0.57\zs \\
      RMSE     &   \z0.66 &     \z1.03 &     \z0.97 &   \z0.72\zs &   \z0.72 &     \z0.60 &     \z0.59 &   \z0.67\zs \\
      SDE      &   \z0.32 &     \z0.94 &     \z0.91 &   \z0.74\zs &   \z0.34 &     \z0.61 &     \z0.60 &   \z0.60\zs \\
      Min      &   \z0.18 &      -2.28 &      -2.14 &    -1.37\zs &   \z0.12 &      -1.40 &      -1.33 &    -1.30\zs \\
      Max      &   \z1.20 &     \z1.29 &     \z1.35 &   \z1.38\zs &   \z1.20 &     \z0.68 &     \z0.70 &   \z1.15\zs \\
    \end{tabular}
  \end{ruledtabular}
\end{table*}
\end{squeezetable}

\begin{figure*}
  \centering
  \includegraphics[width=0.245\linewidth]{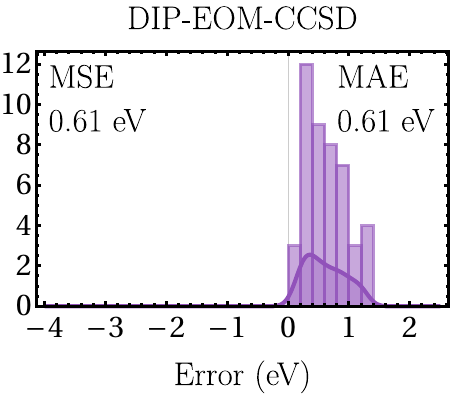}
  \includegraphics[width=0.245\linewidth]{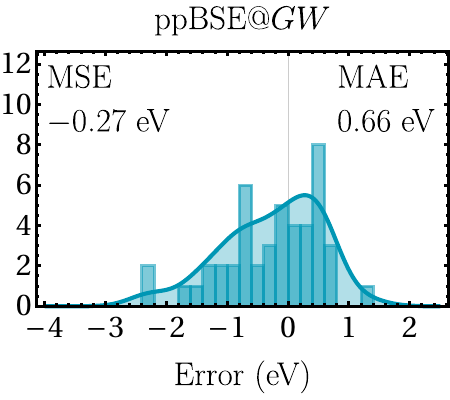}
  \includegraphics[width=0.245\linewidth]{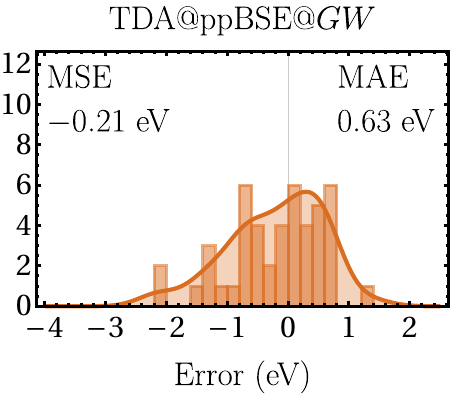}
  \includegraphics[width=0.245\linewidth]{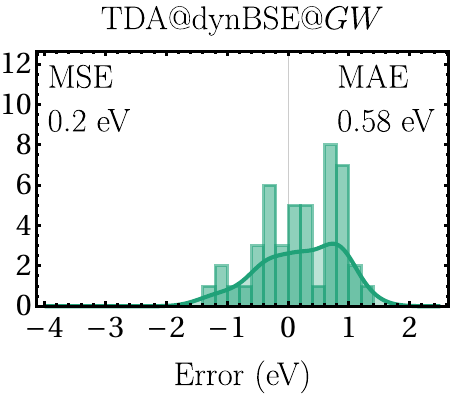}
  \caption{Histogram of the errors (with respect to FCI) for the singlet and triplet principal DIP of 23 small molecules computed in the aug-cc-pVTZ basis set at various levels of theory.
    The leftmost panel gathers the DIP-EOM-CCSD results and the three other panels are based on pp-BSE with a static $GW$ kernel (ppBSE@$GW$), a static $GW$ kernel within the TDA (TDA@ppBSE@$GW$), and a dynamic $GW$ kernel within the TDA (TDA@ppBSE@dynBSE), respectively.}
  \label{fig:error_plot_bse}
\end{figure*}

First, we focus on the $GW$ kernel (see Sec.~\ref{sec:gw_kernel}) which has been the most popular in the eh case.
The results for the three variants of this kernel that will be discussed are reported in Table \ref{tab:tab2}, and the corresponding histograms of errors are displayed in Fig.~\ref{fig:error_plot_bse}.
The static $GW$ kernel, denoted as ppBSE@$GW$, brings a quantitative improvement with respect to both ppRPA@HF and ppRPA@$GW$.
However, the MAE of ppBSE@$GW$, \SI{0.66}{\eV}, is still slightly higher than DIP-EOM-CCSD while having the same $\order{N^6}$ formal computational cost.
The TDA of the static $GW$ kernel (denoted as TDA@ppBSE@$GW$) also has a computational cost of $\order{N^6}$ but with a much lower prefactor as its scaling is formally $\order{O^5V}$ while the full ppBSE$@GW$ is $\order{V^6}$ and DIP-EOM-CCSD is $\order{O^2V^4}$ (where $O$ and $V$ are the numbers of occupied and virtual spin-orbitals, respectively). \cite{Shavitt_2009}
Hence, TDA@ppBSE@$GW$ is much less expensive than ppBSE@$GW$ and DIP-EOM-CCSD.
As one can see in Fig.~\ref{fig:error_plot_bse}, this approximation does not quantitatively affect the MSE and MAE with respect to its ppBSE@$GW$ counterpart.
The impact of the TDA is further investigated using various kernels in the \SupInf.
This shows that the effect (on average) of the TDA, as for the $GW$ kernel, is to increase the DIPs with respect to the full pp-BSE scheme.

The last variant of the $GW$ kernel that is considered is the perturbative correction accounting for dynamic effects.
This correction is only considered in the TDA and the results are referred to as TDA@dynBSE@$GW$ in Table \ref{tab:tab2} and Fig.~\ref{fig:error_plot_bse}.
For the 46 DIPs considered in this work, the dynamic correction is positive for 45 of them.
For the DIPs that were already overestimated at the static level, this positive shift slightly worsened the results.
However, on average, this perturbative correction improves the results as the MAE (\SI{0.58}{\eV}) is decreased with respect to the static case.
Hence, the TDA@dynBSE@$GW$ method has the same accuracy as the EOM-DIP-CCSD while being more centered around zero (MSE of \SI{0.2}{\eV}).
Regarding its cost, computing the dynamic correction scales as $\order*{O^5V}$, but this task has to be performed for each eigenvalue one is willing to correct.
Note that the TDA@dynBSE@$GW$ results have been obtained using $\eta=\SI{0.05}{\hartree}$ in order to regularize the diverging denominators of Eq.~\eqref{eq:gw_dynamic_kernel}.
The renormalization factor associated with the dynamic correction is reported in parenthesis in Table \ref{tab:tab2}.
The smallest factor among this set is 0.78, which shows that, at least under regularization, the perturbation theory is well-behaved.
Finally, the dynamically-corrected $GW$ kernel negates the MAE discrepancy between singlets and triplets that can be observed at the static level.

\begin{squeezetable}
\begin{table}
  \caption{DIPs (in \si{\eV}) toward the singlet (left panel) and triplet (right panel) dication ground states in the aug-cc-pVTZ basis set computed with pp-BSE using a static GF(2) kernel and a static $T$-matrix kernel.}
  \label{tab:tab3}
  \begin{ruledtabular}
    \begin{tabular}{lcccc}
               &    \mc{2}{c}{Singlet DIPs}    &    \mc{2}{c}{Triplet DIPs}    \\
                    \cline{2-3}		           \cline{4-5}
      Molecule & ppBSE@GF(2) & ppBSE@$GT$ & ppBSE@GF(2) & ppBSE@$GT$ \\
      \hline
      \ce{H2O} &       36.54 &      41.51 &       35.90 &      40.93 \\
      \ce{HF}  &       44.76 &      50.96 &       42.12 &      48.71 \\
      \ce{Ne}  &       59.99 &      65.76 &       56.79 &      63.02 \\
      \ce{CH4} &       37.55 &      38.97 &       36.97 &      38.32 \\
      \ce{NH3} &       32.40 &      35.75 &       35.73 &      38.65 \\
      \ce{CO}  &       39.46 &      42.04 &       41.04 &      41.36 \\
      \ce{N2}  &       42.47 &      44.04 &       43.11 &      44.40 \\
      \ce{BF}  &       30.30 &      33.79 &       37.02 &      37.55 \\
      \ce{LiF} &       32.92 &      40.31 &       30.29 &      38.25 \\
      \ce{BeO} &       24.38 &      31.42 &       22.95 &      30.03 \\
      \ce{BN}  &       31.40 &      35.85 &       29.84 &      34.67 \\
      \ce{C2}  &       37.10 &      37.95 &       35.88 &      36.84 \\
      \ce{CS}  &       30.29 &      33.51 &       32.35 &      32.90 \\
      \ce{LiCl}&       28.81 &      30.93 &       27.58 &      29.61 \\
      \ce{F2}  &       42.22 &      44.79 &       40.14 &      44.28 \\
      \ce{H2S} &       29.55 &      31.22 &       31.58 &      32.61 \\
      \ce{PH3} &       28.84 &      30.63 &       31.62 &      32.11 \\
      \ce{HCl} &       35.76 &      37.05 &       34.20 &      35.68 \\
      \ce{Ar}  &       43.07 &      44.78 &       41.56 &      43.15 \\
      \ce{SiH4}&       32.48 &      32.83 &       32.39 &      32.64 \\
      \ce{CH2O}&       30.32 &      32.60 &       33.25 &      35.61 \\
      \ce{CO2} &       37.00 &      38.37 &       35.96 &      37.77 \\
      \ce{BH3} &       36.00 &      36.60 &       34.99 &      35.36 \\
      \hline
      MSE      &       -2.85 &     \z0.11 &       -2.38 &    \z0.28 \\
      MAE      &      \z2.94 &     \z0.45 &      \z2.44 &    \z0.45 \\
      RMSE     &      \z3.58 &     \z0.64 &      \z3.24 &    \z0.61 \\
      SDE      &      \z2.22 &     \z0.65 &      \z2.26 &    \z0.55 \\
      Min      &       -7.80 &      -0.77 &       -7.22 &     -0.92 \\
      Max      &      \z1.12 &     \z1.97 &      \z0.74 &    \z1.71 \\
    \end{tabular}
  \end{ruledtabular}
\end{table}
\end{squeezetable}

\begin{figure}
  \centering
  \includegraphics[width=0.49\linewidth]{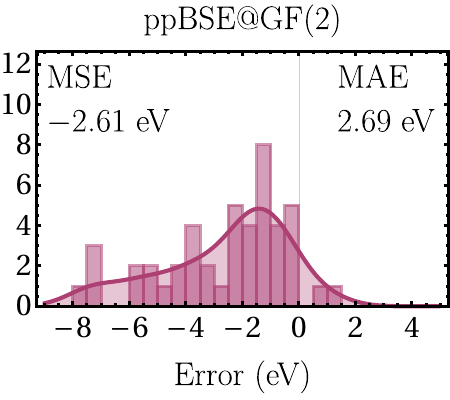}
  \includegraphics[width=0.49\linewidth]{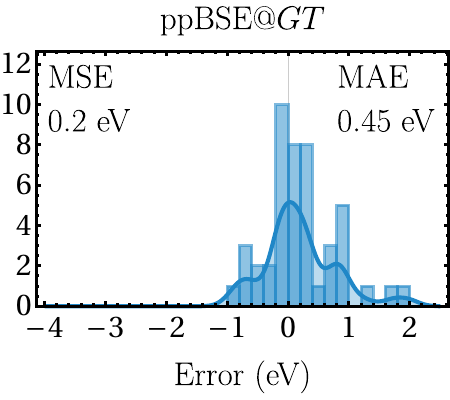}
  \caption{Histogram of the errors (with respect to FCI) for the singlet and triplet principal DIP of 23 small molecules computed using the aug-cc-pVTZ basis set using a static GF(2) kernel and a static $T$-matrix kernel.}
  \label{fig:error_plot_gf2gt}
\end{figure}

To conclude this section, we discuss alternative kernels based on second-order or $T$-matrix self-energies.
The 46 DIPs have been computed using the GF(2) kernel under the static approximation and the results are reported in Table \ref{tab:tab3} under the label ppBSE@GF(2).
As shown by the corresponding histogram of errors (see Fig.~\ref{fig:error_plot_gf2gt}), the static second-order kernel performs as poorly as ppRPA@HF but with a MSE of the opposite sign.
As for the $GW$ case, the effect of the one-body energies and the kernel adds up, leading to a decrease in the DIPs with respect to ppRPA@HF.
However, in the GF(2) case, these two corrections are too large, and, hence, the ppBSE@GF(2) correction is overestimated and does not improve the result on average.

The performance of the second-order $T$-matrix kernel, discussed in Sec.~\ref{sec:gt_kernel}, is also investigated.
This kernel is considered in the static limit and is denoted as ppBSE@$GT$.
The corresponding numerical results are displayed in Table \ref{tab:tab3} and Fig.~\ref{fig:error_plot_gf2gt}.
This method achieves the best MAE (\SI{0.45}{\eV}) among all the methods considered in this work.
Hence, as the $GW$ kernel, the ppBSE@$GT$ also achieves a correct balance between the corrections coming from the one-body energies and the kernel.
On the other hand, this kernel is also more expensive than the $GW$ and GF(2) ones as it requires the computation of the tensor elements associated with the effective interaction $T$ [see Eq.~\eqref{eq:tensor_elem_t}] which formally scales as $\order*{N^4OV}$.
Note also that, in the static $T$-matrix case, there is no noticeable difference in accuracy between singlet and triplet DIPs.

\subsection{Double core hole states}
\label{sec:dch}

DCH states were first discussed by Cederbaum and co-workers, who showed that they are significantly more sensitive to the chemical environment than single-core holes. \cite{Cederbaum_1986a,Cederbaum_1987}
This theoretical prediction was later experimentally confirmed by several research groups. \cite{Lablanquie_2011,Salen_2012,Marchenko_2018,Ismail_2024}
Cederbaum's seminal works have since inspired numerous studies on DCH states using state-specific correlated methods. \cite{Tashiro_2010,Lee_2019,Carniato_2019,Zheng_2020,Ferte_2020,Ferte_2022}
These theoretical developments have been instrumental in accurately interpreting the satellite structure observed in some DCH spectra. \cite{Ferte_2020,Ferte_2022}

The DIPs corresponding to DCH states are naturally captured within the poles of the two-body pp propagator [see Eq.~\eqref{eq:lehman_K}].
Table \ref{tab:tab4} presents single-site DCH energies for two different kernels: the pp-RPA kernel (Sec.~\ref{sec:1st_kernel}), the static $GW$ kernel (Sec.~\ref{sec:gw_kernel}) \titou{and the static $T$-matrix kernel (Sec.~\Ref{sec:gt_kernel}).}
The results are compared with mean-field state-specific ($\Delta$SCF) DCH energies obtained using MOM \titou{and with DIP-EOM-CCSD under the CVS approximation.}
Finally, the reference energies are of FCI quality and have been computed using the CIPSI algorithm under the CVS approximation on top of the corresponding $\Delta$SCF determinant.
In the case of DCHs, the CVS approximation constrains the determinants in the CI \titou{and EOM} expansions to be doubly core ionized configurations.

\begin{figure}
  \centering
  \includegraphics[width=\linewidth]{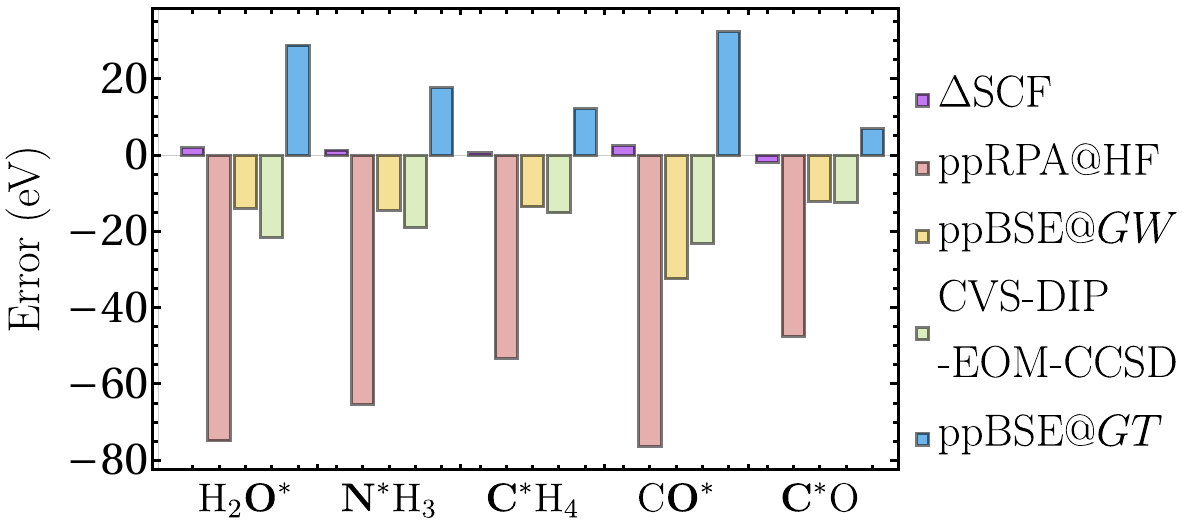}
  \caption{Errors (with respect to CVS-FCI) associated with single-site DCH energies computed in the aug-cc-pCVTZ basis set at the ppRPA@HF, ppBSE@$GW$, and $\Delta$SCF levels.
    The asterisk indicates the ionization site.}
  \label{fig:dch_plot}
\end{figure}

Figure \ref{fig:dch_plot} shows the error of the \titou{five} approximate methods considered in this work.
\titou{It is readily seen that both ppBSE@$GW$ and ppBSE@$GT$ largely improves the DCH energies with respect to ppRPA@HF.}
However, the minimal error for ppBSE@$GW$ remains as large as \SI{12}{\eV}, while the maximum error at the $\Delta$SCF level is only \SI{2.5}{\eV}.
This significant performance disparity arises from the inclusion of orbital relaxation effects at the MOM level, which are absent in the linear-response approach.
\titou{Indeed, the CVS-DIP-EOM-CCSD, which is also a linear-response approach, showcase a similar accuracy as ppBSE@$GW$.
In the CC case, the results would be improved by including triples or higher excitations. \cite{Gururangan_2025}}
This comparison highlights that while kernel improvements enhance results in the linear-response framework, state-specific formalisms should be preferred when available for core-ionized states.
Nevertheless, the improvement offered by ppBSE@$GW$ can be valuable in cases where state-specific approaches are challenging to implement, as in periodic systems. \cite{Aoki_2018,Zhu_2021,Kahk_2021,Kahk_2023}


\begin{table}
  \caption{Single-site DCH energies (in \si{\eV}) computed at various levels of theory with the aug-cc-pCVTZ basis set.
  The asterisk indicates the ionization site.}
  \label{tab:tab4}
  \begin{ruledtabular}
    \begin{tabular}{lccccc}
      Molecule    & H$_2$\textbf{O$^*$} & \textbf{N$^*$}H$_3$ & \textbf{C$^*$}H$_4$ & C\textbf{O$^*$} & \textbf{C$^*$}O \\
      \hline                                                            
      CVS-FCI           & 1172.9 & 892.2 & 651.3 & 1177.0 & 665.9 \\
      $\Delta$SCF       & 1170.9 & 890.9 & 650.8 & 1174.5 & 667.7 \\
      ppRPA@HF          & 1247.9 & 957.5 & 704.6 & 1253.4 & 713.4 \\
      ppBSE@$GW$        & 1186.9 & 906.6 & 664.8 & 1209.4 & 678.0 \\
      CVS-DIP-EOM-CCSD  & 1194.5 & 911.0 & 666.5 & 1200.1 & 678.2 \\
      ppBSE@$GT$        & 1144.1 & 874.5 & 639.2 & 1144.6 & 659.0 \\
    \end{tabular}
  \end{ruledtabular}
\end{table}

\section{Conclusion}
\label{sec:conclusion}

This work provides an in-depth discussion of the eh and pp components of the two-body propagator, highlighting their similarities from the perspectives of both correlation functions and linear response theory. 
In particular, it demonstrates that the response of an anomalous propagator to a pairing field yields a kernel expression for the pp-BSE that is fully analogous to the standard eh-BSE kernel.
This kernel has been explicitly calculated for common self-energy approximations, including the first- and second-order Coulomb self-energies.
Additionally, approximate kernels derived from effective interactions, specifically $W$ (based on bubble diagrams) and $T$ (based on pp-ladder diagrams), are discussed in detail.
These approximate pp kernels hold potential in other contexts, such as the three-body propagator equations, which usually rely on the Coulomb kernel or \textit{ad hoc} kernels. \cite{Deilmann_2016,Riva_2023,Riva_2024}

The performance of these various approximations has been assessed for valence and core DIPs.
The influence of the choice of the starting point and of the kernels has been investigated across a set of 46 DIPs.
It has been shown that the static $GW$ kernel under the TDA is only slightly worse than DIP-EOM-CCSD but with a much lower computational cost.
Furthermore, adding a perturbative correction on top of the pp-BSE brings it even closer to the DIP-EOM-CCSD accuracy.
On the other hand, the second-order Coulomb kernel performs poorly for DIPs.
However, the second-order kernel relying on the pp $T$-matrix effective interaction is more accurate than ppBSE@$GW$ and DIP-EOM-CCSD on average, albeit with a higher computational cost.
Finally, it has been shown that the $GW$ kernel also provides a quantitative improvement over pp-RPA for core DIPs.

While this work has focused on DIPs, the accuracy of the various kernels for DEAs is an obvious and interesting follow-up of this work.
However, systems capable of binding two electrons are generally spatially large, posing challenges for our current computational implementations.
In addition, the DEAs of the ($N-2$)-electron system computed at the pp-RPA level have been extensively used to compute neutral excitation energies of the $N$-electron system. \cite{Yang_2013b,Yang_2014a,Yang_2015,Yang_2016,Yang_2017a,Li_2024a,Li_2024b}
Comparing the performance of various kernels within this framework would be valuable.
Finally, the adiabatic connection fluctuation dissipation theorem, which has been applied to the two-body eh propagator to compute correlation energies, could be transposed to the pp case as well. \cite{Furche_2005,vanAggelen_2013,vanAggelen_2014,Holzer_2018b,Loos_2020e}
This is left for future work.

\section*{Supplementary Material}
\label{sec:supmat}

See the \SupInf for a detailed derivation of every equation presented in the main manuscript and additional results on the pp-RPA starting-point dependence and the TDA.

\acknowledgments{
  \titou{The authors thank Karthik Gururangan and Piotr Piecuch for their careful checking of the data and for bringing to our attention the CVS-DIP-EOM-CCSD implementation of \textsc{ccpy}.
  The authors thank Robert van Leeuwen} and Noam Mandin for fruitful discussions at the early stage of this project.
This project has received funding from the European Research Council (ERC) under the European Union's Horizon 2020 research and innovation programme (Grant agreement No.~863481).}

\section*{Data availability statement}
The data that supports the findings of this study are available within the article and its supplementary material.

\bibliography{anom_propag}

\end{document}